\documentclass[reprint, 
nofootinbib,
amsmath,amssymb,
aps,prd,
]{revtex4-2}

\usepackage{graphicx}
\usepackage{bm}
\usepackage{booktabs}
\usepackage{amsmath}
\usepackage{mathrsfs}
\usepackage{subfigure}
\usepackage{enumerate}

\usepackage[hidelinks]{hyperref}
\hypersetup{
    colorlinks=true,
    linkcolor=blue,
    filecolor=gray,      
    urlcolor=blue,
    citecolor=blue,
}

\allowdisplaybreaks[1]
            
\begin{document}


\title{Post-Newtonian dynamics of charged compact binaries}

\author{Zi-Han Zhang$^{1,2}$}
 \email{zhangzihan242@mails.ucas.ac.cn}

\author{Tan Liu$^{3,4,5}$}
 \email{lewton@mail.ustc.edu.cn}

\author{Shuai Zhang$^{6}$}
 \email{zhangshuai21@lzu.edu.cn}
 
\author{Zong-Kuan Guo$^{4,5,7}$}
\email{guozk@itp.ac.cn}

\affiliation{$^{1}$International Centre for Theoretical Physics Asia-Pacific, University of Chinese Academy of Sciences, 100190 Beijing, China}
\affiliation{$^{2}$Taiji Laboratory for GW Universe, University of Chinese Academy of Sciences, 100049 Beijing, China}

\affiliation{$^{3}$School of Physics and Optoelectronic Engineering, Yangtze University, Jingzhou 434023, China}

\affiliation{$^{4}$School of Fundamental Physics and Mathematical Sciences, Hangzhou Institute for Advanced Study, University of Chinese Academy of Sciences, Hangzhou 310024, China}

\affiliation{$^{5}$School of Physical Sciences, University of Chinese Academy of Sciences, Beijing 100049, China }

\affiliation{$^{6}$School of Physical Science and Technology, Lanzhou University, Lanzhou 730000, China }

\affiliation{$^{7}$Institute of Theoretical Physics, Chinese Academy of Sciences, Beijing 100190, China }


\date{\today}

\begin{abstract}
We investigate the dissipative dynamics of charged compact binaries in Einstein-Maxwell theory. By evaluating the mass and electric multipole moments, we compute the gravitational and electromagnetic fluxes  {through next-to-leading order in the post-Newtonian expansion}. Using the flux-balance equations, we derive the evolution of the orbital angular frequency for quasi-circular inspirals. We further analyze circular orbit stability in charged black-hole binaries and quantify how the charge-to-mass ratios affect the inspiral dynamics.

\end{abstract}

\maketitle
 
\section{Introduction}

Hundreds of compact binary systems have been observed  
by LIGO, Virgo, and KAGRA \cite{PhysRevLett.116.061102,PhysRevLett.116.241103,PhysRevLett.118.221101,PhysRevLett.119.141101,PhysRevLett.125.101102}. General relativity admits the Kerr-Newman family as the stationary, asymptotically flat electrovacuum black hole (BH) solution, characterized by mass, spin, and charge \cite{PhysRevLett.11.237,10.10631.1704351}. Kerr-like BHs in astronomical environments might possess small charge-to-mass ratios \cite{10.1093/mnras/sty2182,Zajaček_2019,PhysRevD.109.024058}, although standard plasma processes are expected to neutralize them on short timescales \cite{Gibbons1975,10.1093/mnras/stz1904,Cardoso_2016}. Nevertheless, charged black holes remain useful probes of early-Universe scenarios and gravitational wave (GW) phenomenology, particularly in the context of primordial black holes (PBHs) \cite{PhysRevLett.117.061101,PhysRevD.96.123523,Chen_2018}. It is now possible to extract observational constraints on the charges of BHs through GW analyses, which motivates a detailed investigation of dissipative processes in charged binary systems.

Charged binaries involve both electromagnetic (EM) and gravitational interactions, leading to a richer phenomenology than neutral binaries. Previous studies have examined Einstein-Maxwell theory in the post-Newtonian (PN) framework and calculated the conservative dynamics through 2PN order \cite{PhysRevD.108.024020,PhysRevD.109.084048,placidi2025chargedblackholebinaryevolution, PhysRevD.98.104010,tang2025inspiralingbinarychargedblack,verma2026postnewtoniandynamicsradiatingcharges,alonzoartiles2026reissnernordstromblackholessecond}. For binaries with large charge-to-mass ratios, both the leading-order Coulomb interaction and EM radiation can significantly affect the dynamics. Accurate PN trajectories are needed to describe strongly radiating charged BBHs. In addition, the PN approach extends to modeling white dwarf-neutron star binaries \cite{PhysRevLett.119.161101}.

The main purpose of this work is to analyze how EM and GW radiation affect the frequency evolution of charged binary systems. This is directly motivated by the need for accurate GW waveform templates that account for possible charge effects in BBHs and other compact objects, which may be constrained by  GW observations. Specifically,  {we rederive the  Fokker action in the framework of Einstein-Maxwell theory, both to fix the notation and to provide a more detailed derivation. }We check the result against the Reissner-Nordström (RN) limit and obtain the 1PN Lagrangian with charges, which is consistent with  previous works \cite{placidi2025chargedblackholebinaryevolution, PhysRevD.98.104010}. We obtain the Noether charges and use them to transform from individual-particle coordinates to the center-of-mass frame.
We employ the flux-balance equation to incorporate radiation back-reaction. The GW and EM fluxes are derived through next-to-leading order in the PN expansion, with partial results obtained for the EM flux at next-to-next-to-leading order. Because the leading order EM flux enters at -1PN relative to the leading order GW flux, the total -1PN and 0PN fluxes are complete, whereas the 1PN flux is incomplete. The leading EM back-reaction effect was studied in
\cite{placidi2025chargedblackholebinaryevolution}; our results extend those of \cite{placidi2025chargedblackholebinaryevolution} by one PN order.
Solving the angular-momentum flux-balance equation for quasi-circular orbits, we determine the evolution of the gauge-invariant orbital angular frequency. Further, to deepen our understanding of charged binary dynamics, we analyze the stability of circular orbits and determine the innermost stable circular orbit (ISCO). We numerically compute the GW frequency as a function of time and plot the energy radiation power and the orbital trajectories for systems with different charge-to-mass ratios and different charge signs.


This paper is organized as follows. In Sec. \ref{Sec2}, we derive the 1PN metric and Fokker action of Einstein–Maxwell theory for charged binaries. In Sec. \ref{Sec3}, we derive the equation of motion and acceleration at 1PN order in harmonic gauge. By analyzing Noether charges, we transform the acceleration to the center-of-mass frame. In Sec. \ref{Sec4},  we calculate the orbital angular frequency evolution driven by EM and GW radiation and analyze the stability of circular orbits for charged BBHs. We conclude in Sec. \ref{Sec5} Throughout this work, $G$ is the gravitational constant, $c$ is the speed of light.

\section{Effective action in Einstein-Maxwell theory}\label{Sec2}

 {In this section, we rederive the 1PN results in Einstein-Maxwell theory, both to fix the notation for the calculations and to provide a more detailed derivation.}
\subsection{General construction}
We consider a binary point-particle system with two dynamical fields: a metric $g_{\mu\nu}$ and the EM four-potential $A_{\mu}$ \cite{PhysRevD.108.024020,placidi2025chargedblackholebinaryevolution,tang2025inspiralingbinarychargedblack}. The total action is expressed as
\begin{equation}\label{action}
S=S_g[g]+S_{EM}[g,A]+S_I[g,A]+S_m,
\end{equation}
where $S_g$ is the action of the gravitational field, $S_{EM}$ is the action of the EM field, $S_I$ is the interaction part of the EM field, and $S_m$ is the action of matter. The gravitational action is
\begin{equation}\label{Gaction}
S_g=\frac{c^3}{16\pi G}\int\mathrm{d}^4x\sqrt{-g}\Big(R-\frac{1}{2}g_{\mu\nu}\Gamma^\mu\Gamma^\nu\Big),
\end{equation}
where $g=\text{det}(g_{\mu\nu})$, $R$ is the Ricci scalar, $g_{\mu\nu}\Gamma^\mu\Gamma^\nu$ is the gauge fixing term,  $\Gamma^\alpha\equiv g^{\mu\nu}\Gamma^{\alpha}_{\mu\nu}$, and 
$\Gamma_{\mu\nu}^\alpha$ denotes the Christoffel symbol. We define the dimensionless vector field $A_\mu=\sqrt{G /(4\pi\epsilon_0 c^4)} \hat{A}_{\mu}$, where $\hat{A}_{\mu}$ is the physical EM field and the vacuum permittivity satisfies $\epsilon_0 = 1/(4\pi)$. The action of the EM field is
\begin{equation}
S_{EM}=-\frac{c}{16\pi  }\int\mathrm{d}^4x\sqrt{-g}\Big(F_{\mu\nu}F^{\mu\nu}+2(\nabla_\mu A^\mu)^2\Big),
\end{equation}
where $(\nabla_\mu A^\mu)^2$ is the gauge-fixing term in the Lorenz gauge $\nabla_\mu A^\mu=0$, $F_{\mu\nu}=\nabla_\mu A_{\nu}-\nabla_\nu A_{\mu}=2\partial_{[\mu}A_{\nu]}$ is the antisymmetric field-strength tensor, whose components define the electric and magnetic fields as
\begin{equation}
E_i=-cF_{0i},\quad B_i=\frac{1}{2}\epsilon_{ijk}F_{jk}=\epsilon_{ijk}\partial_j A_k.
\end{equation}
The coupling between the EM field and the   { current $j^\mu_A$ is
\begin{equation}
S_I=\sum_{A=1,2} \int \mathrm{d}^4x \sqrt{-g} j_A^\mu A_\mu,
\end{equation}
where subscripts $A=1,2$ denote the index of the binary components. The current for a point particle charge A is
\begin{equation}
    j_A^\mu\equiv \frac{\delta_A}{\sqrt{-g}} q_A v^\mu_A,
\end{equation}
which satisfies $\nabla_\mu j^\mu_A=0$. $q_A$ is the electric charge and $v_A^\mu=\mathrm{d}x_A^\mu/\mathrm{d}t$ is the coordinate-time velocity of particle A.} The matter action of a point-particle system is
\begin{equation}
S_m=-\sum_{A=1,2} m_Ac^2\int \mathrm{d}^4x\delta_A\sqrt{-g_{\mu\nu}v_A^\mu v_A^\nu/c^2},
\end{equation}
where $\delta_A\equiv \delta^{(3)}(\bm{x}-\bm{x}_A(t))$ is the three-dimensional Dirac delta distribution and $\bm{x}_A(t)$ is the worldline of particle A.

We now expand the action in the PN framework using the counting $\epsilon\thicksim v^2/c^2$, where $v$ represents the characteristic orbital velocity. We call a quantity of order $\epsilon^a\thicksim \mathcal{O}(c^{-2a})$ an aPN contribution.

\subsection{Field equations and metric}
Varying the action with respect to $g_{\mu\nu}$ and $A_\mu$ while retaining terms through 1PN order, we obtain \cite{PhysRevD.108.024020}
\begin{align}
    \Box h^{\mu\nu}&=\frac{16\pi G}{c^4}(T^{\mu\nu}_{\text{pp}}+\Lambda^{\mu\nu}),\label{Fieldh}\\
    \Box A_{\mu}&=-\frac{4\pi  }{c^2} \iota_\mu,\label{FieldEQ}
\end{align}
  {where the electromagnetic source is
\begin{equation}\label{iota}
    \iota_{A}^\mu=\sqrt{-g}j^\mu_A-\frac{c^2}{4\pi}\Phi_A^\mu.
\end{equation}}
  {The stress-energy tensor of compact point-particles  is \cite{landau2013classical,Blanchet2024}
\begin{equation}\label{Tmunu}
    T^{\mu\nu}_{\text{pp}}=\sum_{A=1,2}\frac{m_A}{\sqrt{-g}}  \frac{\delta_A v^\mu_A v^\nu_A}{\sqrt{-g_{\alpha\beta} v_A^\alpha v_A^\beta/c^2}},
\end{equation}}
The non-compact stress-energy pseudotensor of gravity and the EM field $\Lambda^{\mu\nu}=\Lambda^{\mu\nu}_g+\Lambda^{\mu\nu}_{EM}$ is \cite{PhysRevD.108.024020}
\begin{align}
    \Lambda^{\mu\nu}_g=&-h^{\alpha\beta}\partial_\alpha\partial_\beta h^{\mu\nu}+\frac{1}{2}\partial^{\mu}h_{\alpha\beta}\partial^{\nu}h^{\alpha\beta}-\frac{1}{4}\partial^{\mu}h\partial^{\nu}h\notag\\
    &+\partial_{\alpha}h^{\mu\beta}\big(\partial^\alpha h^{\nu}_{\beta}+\partial_\beta h^{\nu\alpha}\big)-2\partial^{(\mu} h_{\alpha\beta} \partial^\alpha h^{\nu)\beta}\notag\\
    & -\frac{1}{4}\eta^{\mu\nu}\partial_\tau h_{\alpha\beta} \partial^\tau h^{\alpha\beta} + \frac{1}{8}\eta^{\mu\nu}\partial_\alpha h \partial^\alpha h\notag\\
    &+ \frac{1}{2}\eta^{\mu\nu}\partial_\alpha h_{\beta\tau} \partial^\beta h^{\alpha\tau} +\mathcal{O}(h^3),\\
    \Lambda^{\mu\nu}_{EM}=&\frac{|g| c^2}{4\pi}\big(F^{\mu\lambda}F^\nu_{\lambda}-\frac{1}{4}g^{\mu\nu}F_{\alpha\beta}F^{\alpha\beta}\big).
\end{align}
The noncompact nonlinear EM source term is \cite{PhysRevD.108.024020}
\begin{align}\label{Phi}
    \Phi_\mu=&- h^{\alpha\beta} \partial_\alpha\partial_\beta A_\mu - \partial_\mu h^{\alpha\beta} \partial_\alpha A_\beta - F_{\alpha\beta} \partial^\alpha h_\mu^\beta\notag\\
    &- \frac{1}{2} F_{\mu\alpha} \partial^\alpha h + \mathcal{O}(h^2 A),
\end{align}
where $h=h^\alpha_\alpha$. The EM vector field is
\begin{align}\label{Amu}
    A_0&=\frac{1}{c}\varphi-\frac{1}{c^3}V\varphi+\mathcal{O}(c^{-5}),\\
    A_i&=\frac{1}{c^2}\chi_i+\mathcal{O}(c^{-4}).
\end{align}
where $\varphi$ and $ \chi_i$ are the EM potentials. Treating $\Lambda^{\mu\nu}$ as a perturbation source as $T^{\mu\nu}$ in Eq. \eqref{Fieldh} and gives the gravitational-field amplitude $h^{\mu\nu}$ as
\begin{align}
    h^{00}&= -\frac{4}{c^2}V+\frac{1}{c^4}\big(-8V^2-2W^i_i+G\varphi^2\big)\notag\\
    &+\mathcal{O}(c^{-6}),\\
    h^{0i}&=-\frac{4}{c^3}V^i+\mathcal{O}(c^{-5}),\\
    h^{ij}&=-\frac{2}{c^4}\big(2W^{ij}-2\delta^{ij}W_k^k+2\phi^{ij}-\delta^{ij}\varphi^2\big)\notag\\
    &+\mathcal{O}(c^{-6}),
\end{align}
where $V,V_i$ and $W$ are post-Newtonian potentials \cite{Blanchet2024}. $\phi^{ij}$ is another electric potential. Using the relationship between the gravitational-field amplitude $h^{\mu\nu}$ and the contravariant metric $g^{\mu\nu}$
\begin{equation}
    h^{\mu\nu}\equiv\sqrt{-g}g^{\mu\nu}-\eta^{\mu\nu},
\end{equation}
we obtain the 1PN covariant metric 
\begin{align}
    g_{00}&=-1+\frac{2}{c^2}V-\frac{1}{c^4}\big(2V^2+G\varphi^2\big)+\mathcal{O}(c^{-6}),\\
    g_{0i}&=-\frac{4}{c^3}V_i+\mathcal{O}(c^{-5}),\\
    g_{ij}&=\delta_{ij}\big(1+\frac{2}{c^2}V\big)+\mathcal{O}(c^{-4}).
\end{align}
It is instructive to compare the following calculation with the Reissner‑Nordström metric
\begin{align}\label{RN}
    \mathrm{d}s^2=&-c^2f(r)\mathrm{d}t^2+f(r)^{-1}\mathrm{d}r^2+r^2\mathrm{d}\Omega,
\end{align}
where $\mathrm{d}\Omega=\mathrm{d}\theta^2+\sin^2\theta \mathrm{d}\phi^2$ and 
\begin{equation}
        f(r)=1-\frac{2 G M}{c^2 r}+\frac{Q^2}{r^2}
\end{equation}
with  $Q\equiv \sqrt{G  /c^4}q$. This definition helps us to match the leading order  EM effect to the 1PN order in the PN-EM metric and the EM field in Eq. \eqref{Amu}. 

The PN potentials in the metric are expressed as
\begin{align}
    \quad\qquad\qquad\qquad\Box& V&=&-4\pi G\sigma,&\qquad\qquad\\
    \Box& V_i&=&-4\pi G\sigma_i,&\\
    \Box& \varphi&=&-4\pi G  c^2 \iota_0,&\\
    \Box& \chi_i&=&- 4\pi G  c^3 \iota_{i}.&
\end{align}
The 1PN potentials $\phi_{ij}$ and $W_{ij}$ are expressed in terms of the potential $V$
\begin{align}
\Box\phi_{ij}&=\partial_i \varphi\partial_j \varphi,\\
    \Box W_{ij}&=-4\pi G(\sigma_{ij}-\delta_{ij}\sigma_{k}^{\;\;k})-\partial_i V\partial_j V,
\end{align}
where $\sigma={c^{-2}} (T^{0}_{0}+T^{i}_{i}), \sigma_i={c^{-1}}T_{0i}, \sigma^{ij}=T^{ij}$ denote the compact-support parts of the source. In particular, we have $\Box\phi_{ii}=\Box \varphi^2+\mathcal{O}(c^{-2})$. The PN potential $V$ is related to the Newtonian potential $U$ by $\Box V=(\bigtriangleup-c^{-2}\partial_t^2)V=\bigtriangleup U =-4\pi G\sigma$, and $\bigtriangleup\equiv\partial_i\partial^i$. 

Using Hadamard partie-finie regularization \cite{Blanchet_2000}, as detailed in Appendix \ref{App.A}, we obtain gravitational potentials
\begin{align}\label{PNpotentials}
    V|_1&=\frac{Gm_2}{r}\bigg\{1+\frac{1}{c^2}\bigg[\frac{3}{2}v_1^2-\frac{Gm_1}{r}+\frac{1}{2}\bm{v}_1\cdot\bm{v}_2\notag\\
    &+\frac{1}{2}(\bm{v}_1\cdot\bm{n})(\bm{v}_{2}\cdot\bm{n})\bigg]\bigg\}+\mathcal{O}(c^{-4}),\\
    V_i|_{1}&=\frac{Gm_2}{r}v_{2i}+\mathcal{O}(c^{-2}),\\
    \varphi|_1&=\frac{ q_2}{r}+\frac{1}{c^2}\frac{G}{r^2}(m_2q_1-m_1q_2)+\mathcal{O}(c^{-4}),\\
    \chi_i|_1&=\frac{ q_2}{2r}\bigg[-(\bm{v}_2\cdot\bm{n})n_{i}-v_{2i}\bigg]+\mathcal{O}(c^{-2}),
\end{align}
where subscripts $1,2$ denote the binary component. $r=|\bm{r}_1-\bm{r}_2|$ and $\bm{n}=(\bm{r}_1-\bm{r}_2)/|\bm{r}_1-\bm{r}_2|$. The symbol $|_1$ denotes the field at position $\bm{x}_1$. The potentials at position  $\bm{x}_2$ can be obtained by interchanging the indices of the binary components in the above expressions, i.e., $1\leftrightarrow 2$.   {The
motion of the charges sources the magnetic potential $\chi_i$.}
\\
\subsection{The Fokker action}
Consider the EM field equation \eqref{FieldEQ}.   {Substituting $\nabla^\mu F_{\mu\nu}=-4\pi G c^{-2}j_{\nu}$} into the action of the EM field, we obtain the effective action of the EM interaction 
\begin{equation}
    S_{EM-eff}=S_{EM}+S_{I}=\frac{1}{2}\sum_{A=1,2} \int \mathrm{d}^4x\sqrt{-g} j_A^\mu A_\mu.
\end{equation}
This effective action $S_{EM-eff}$   {avoids double counting of the EM effects between $S_{EM}$ and $S_{I}$. }The Fokker action for the gravitational field and matter has the usual form, supplemented by terms involving the EM potential $\varphi$, which arises from the stress-energy tensor
\begin{align}
    S_{g}&=\int\mathrm{d}t\sum_{A=1,2} m_A\bigg\{\frac{v_A^2}{2}-\frac{V|_A}{2}\notag\\
    &+\frac{1}{c^2}\bigg(-\frac{3}{4}V|_Av_A^2+2v_A^i V_i|_A+\frac{1}{2}V|_A^2-\frac{1}{2}\varphi|_A^2\bigg)\bigg\}\notag\\[6pt]
    &+\mathcal{O}(c^{-4}),\label{Sg}\\
    S_{m}&=\int\mathrm{d}t\sum_{A=1,2} m_A\bigg\{-c^2+\frac{v_A^2}{2}+V|_A\notag\\
    &+\frac{1}{c^2}\bigg(\frac{3}{2}V|_Av_A^2+\frac{v_A^4}{8}-4v_A^i V_i|_A-\frac{1}{2}V|_A^2+\frac{1}{2}\varphi|_A^2\bigg)\bigg\}\notag\\[6pt]
    &+\mathcal{O}(c^{-4}),\label{Sm}
\end{align}
which agrees with the standard Fokker action in \cite{Damour1985, PhysRevD.93.084037} and involves additional EM terms arising from the effective stress-energy tensor in Eq. \eqref{tau} in Appendix \ref{App.B}. 

The total Lagrangian of the binary system is
\begin{equation}\label{Lagrangian}
    L=L_{EM}+L_g+L_k,
\end{equation}
where $L_{EM},L_{g}$ and $L_{k}$ are the Lagrangians of the effective EM interaction, gravitational field,   { and kinetic Lagrangian of the point
particles}, respectively \cite{p1y1-1kcr}.
\begin{widetext}
\begin{align}
    L_{EM}=&-\frac{q_1q_2 }{r}+\frac{1}{c^2}\bigg\{\frac{q_1q_2}{r} \bigg[\frac{G m}{r}+\frac{1}{2}(\bm{v}_1\cdot\bm{n})(\bm{v}_{2}\cdot\bm{n})+\frac{1}{2}\bm{v}_1\cdot\bm{v}_2\bigg]-\frac{1}{2}\frac{G}{r^2}\bigg[m_2 q_1^2+m_1 q_2^2\bigg]\bigg\}+\mathcal{O}(c^{-4}),\\
L_g=&\frac{Gm_1m_2}{r}+\frac{1}{c^2}\frac{Gm_1m_2}{r}\bigg\{\frac{3}{2}v_1^2+\frac{3}{2}v_2^2-\frac{7}{2}\bm{v}_1\cdot\bm{v}_2-\frac{1}{2}(\bm{v}_1\cdot\bm{n})(\bm{v}_2\cdot\bm{n}1)-\frac{1}{2}\frac{Gm}{r}\bigg\}+\mathcal{O}(c^{-4}),\\
L_k=&\frac{1}{2}m_1v_1^2+\frac{1}{2}m_2v_2^2+\frac{1}{c^2}\bigg\{\frac{1}{8}m_1 v_1^4+\frac{1}{8}m_2 v_2^4\bigg\}+\mathcal{O}(c^{-4}),
\end{align}
\end{widetext}
where $\bm{n}=\bm{n}_1-\bm{n}_2$ and the total mass of the binary $m=m_1+m_2$. This 1PN Lagrangian is consistent with Refs. \cite{placidi2025chargedblackholebinaryevolution,PhysRevD.98.104010} and   {the pure electromagnetic part of $L_{EM}$ agrees with Darwin potential }\cite{Jackson:1998nia,POLONYI2014239}. Direct EM interactions are proportional to $ q_1 q_2$, 
and the 1PN terms of $A_\mu$  contribute to the terms of $G m  q_1 q_2/c^2r^2$, $G m_1 q_2^2/c^2r^2$ and $G m_2 q_1^2/c^2r^2$.
  {The terms proportional to $q_1^2$ or $q_2^2$ arise from the gravitational field sourced by the EM energy, which agrees with the  RN metric. }

\section{Equations of motion}\label{Sec3}
In this section, we compute the acceleration of a point-particle binary with charges in the center-of-mass frame, and discuss the Noether charges at 1PN order.
\subsection{The 1PN acceleration in harmonic coordinates}

We first compute the conservative dynamics through 1PN order. Varying the action with respect to $x_1^i$ gives the Euler-Lagrange equation 
\begin{equation}\label{E-L}
    \frac{\partial {L}}{\partial x_{A}^i}-\frac{\mathrm{d}}{\mathrm{d}t}\left(\frac{\partial {L}}{\partial v_A^i}\right)=0.
\end{equation}
We then obtain the acceleration $\bm{a}_A$ as
\begin{align}\label{aA}
    \bm{a}_1=&-\frac{Gm_2}{r^2}\Big[\left(1+\bar{\mathcal{A}}_g\right)\bm{n}+\bar{\mathcal{B}}_g\bm{v}\Big]+\bar{\mathcal{C}}_{g-EM}\bm{n}\notag\\
    &+\frac{1}{r^2}\frac{q_1q_2}{m_1}\Big[(1+\bar{\mathcal{A}}_{EM})\bm{n}+\bar{\mathcal{B}}_{EM}\bm{v}\Big]+\mathcal{O}(c^{-4}),
\end{align}
where  $\bm{v}=\bm{v}_1-\bm{v}_2$ and
\begin{align}
    \bar{\mathcal{A}}_g&=-5\frac{Gm_1}{r}-4\frac{Gm_2}{r}-\frac{3}{2}(\bm{v}_2\cdot\bm{n})^2\notag\\
    &-4\bm{v}_A\cdot\bm{v}_2+v_1^2+2v_2^2,\\
    \bar{\mathcal{A}}_{EM}&=\frac{ q_1q_2}{m_2 r}-7\frac{Gm_1}{r}-5\frac{Gm_2}{r}-\frac{3}{2}(\bm{v}_2\cdot\bm{n})^2\notag\\
    &-\bm{v}_1\cdot\bm{v}_2-\frac{1}{2}v_1^2+\frac{1}{2}v_2^2,\\[6pt]
    \bar{\mathcal{B}}_g&=-4(\bm{v}_1\cdot\bm{n})+3(\bm{v}_2\cdot\bm{n}),\\[10pt]
    \bar{\mathcal{B}}_{EM}&=-(\bm{v}_1\cdot\bm{n}),\\[2pt]
    \bar{\mathcal{C}}_{g-EM}&=\frac{G}{r^3}\Big(\frac{m_2}{m_1}q_1^2+q_2^2\Big).
\end{align}
The coefficient $\bar{\mathcal{C}}_{g-EM}$   {arises from the
gravitational field sourced by the electromagnetic energy}.

\subsection{Noether charges}
The 1PN Fokker action is invariant under the Poincaré group. There are 10 Noether charges corresponding to 10 transformations: the Hamiltonian $\mathcal{H}$, the linear and angular momentum $\mathcal{P}^i$ and $\mathcal{J}^i$, and the center-of-mass integral $\mathcal{N}^i$. The Noether charges are conserved by the  1PN dynamics; radiation, beginning with EM dipole emission at 1.5PN order, breaks these conservation laws.

Consider an infinitesimal deformation of the paths of the two particles $\delta x_A^i(t) \equiv {x'}_A^{i}(t)-x_A^i(t)$. The corresponding linearized perturbation of the Lagrangian is given by \cite{LucBlanchet_2003}.
\begin{equation}\label{deltaL}
    \delta L=\sum_{A=1,2} \bigg\{\frac{\mathrm{d}}{\mathrm{d}t}\big(p_A^i\delta x^i_A\big)+\bigg[\frac{\partial L}{\partial x^i_A}-\frac{\mathrm{d}}{\mathrm{d}t}p_A^i\bigg]\delta x^i_A\bigg\}+\mathcal{O}(\delta x_A^2).
\end{equation}
where 
\begin{equation}
    p_A^i=\frac{\partial L}{\partial v^i_A}.
\end{equation}
For the Lorentz boost, the change in the position of particle A to linear order in the boost velocity $\mathcal{V}^i$ is \cite{LucBlanchet_2003}
\begin{equation}
    \delta x^i_A=-\mathcal{V}^i t+\frac{1}{c^2}\mathcal{V}^jx^j_A v^i_A+\mathcal{O}(\mathcal{V}^2).
\end{equation}
The variation of Lagrangian $\delta L$ is a total time derivative, which can be simply expressed as
\begin{equation}
    \delta L=\mathcal{V}^i\frac{\mathrm{d}Z^i}{\mathrm{d}t}+\mathcal{O}(\mathcal{V}^2).
\end{equation}
The conserved boost integral is $K^i=\mathcal{N}^i-\mathcal{P}^i t$, where $\mathcal{P}^i$ is the total linear momentum, and $\mathcal{N}^i$ is the center-of-mass integral. These quantities are given by
\begin{align}
\mathcal{H}=&\sum_{A=1,2} p^i_Av^i_A-L,\label{HH}\\
    \mathcal{P}^i=&\sum_{A=1,2} p^i_A,\\
    \mathcal{J}^i=&\epsilon_{ijk}\sum_{A=1,2} x^j_A p^k_A,\label{JJ}\\
    \mathcal{N}^i=&-Z^i+\sum_{A=1,2}\frac{1}{c^2}x^i_Ap^k_Av_A^k.\label{PG}
\end{align}
  {Note that there is no acceleration in the 1PN Lagrangian, which appear at 2PN order in the harmonic coordinates. }
The conservation of linear momentum $\mathrm{d}\mathcal{P}^i/\mathrm{d}t=0$ and $\mathrm{d}K^i/\mathrm{d}t=0$ imply $\mathrm{d}\mathcal{N}^i/\mathrm{d}t=P^i$.   {We work in the center-of-mass frame,
where $P^i=0$.} Substituting Lagrangian Eq. \eqref{Lagrangian} into Eq. \eqref{deltaL} and Eq. \eqref{PG} yields
\begin{align}
    \mathcal{N}^i&=m_1x_1^i+\frac{1}{c^2}\bigg\{\frac{1}{2}m_1v_1^2-\frac{G m_1m_2}{2r}+\frac{ q_1q _2}{2r}\bigg\}x_1^i\notag\\[6pt]
    &+(1\leftrightarrow 2)+\mathcal{O}(c^{-4}).
\end{align}
The relative separation, velocity, and acceleration of the binary are
\begin{gather}
    x^i=x^i_1-x^i_2,\quad r=|x^i|,\quad n^i=\frac{x^i}{r}\notag\\
    v^i=v^i_1-v^i_2,\quad a^i=a^i_1-a^i_2.
\end{gather}

The positions and velocities of two particles in the center of mass frame are therefore given by 
\begin{align}
    x^i_1=&+\chi_2 x^i+\frac{1}{c^2}\frac{1}{2}\eta(\chi_1-\chi_2)\Big(v^2-\frac{Gm}{r}\mathcal{Z}\Big) x^i\notag\\
    &+\mathcal{O}(c^{-4}),\label{xCoM}\\
    x^i_2=&-\chi_1 x^i+\frac{1}{c^2}\frac{1}{2}\eta(\chi_1-\chi_2)\Big(v^2-\frac{Gm}{r}\mathcal{Z}\Big) x^i\notag\\
    &+\mathcal{O}(c^{-4}),\\
    \quad v^i_1=&+\chi_2 v^i+\frac{1}{c^2}\bigg\{\frac{1}{2}\eta(\chi_1-\chi_2)\Big(v^2-\frac{Gm}{r}\mathcal{Z}\Big) v^i\notag\\
    &+\frac{Gm}{r^2}\eta\mathcal{Z}(\bm{v}\cdot\bm{n})x^i\bigg\}+\mathcal{O}(c^{-4}),\\
    v^i_2=&-\chi_1 v^i+\frac{1}{c^2}\bigg\{\frac{1}{2}\eta(\chi_1-\chi_2)\Big(v^2-\frac{Gm}{r}\mathcal{Z}\Big) v^i\notag\\
    &+\frac{Gm}{r^2}\eta\mathcal{Z}(\bm{v}\cdot\bm{n})x^i\bigg\}+\mathcal{O}(c^{-4}),
\end{align}
where $\chi_A=m_A/m$, $\eta=m_1m_2/m^2$ is the symmetric mass ratio, $\mathcal{Q}= q_1q_2/Gm_1m_2$ is the symmetric charge to mass ratio. For compactness, we define a gravitational and electric coupling coefficient $\mathcal{Z}=1-\mathcal{Q}$, which combines Newtonian gravitational and electric interactions in a compact form.   {The terms proportional to $\mathcal{Q}$ come from the effects of the EM interactions. }

Now we proceed to the center-of-mass frame. Returning to the energy and the angular momentum in Eq. \eqref{HH} and Eq. \eqref{JJ}, we obtain
\begin{widetext}
\begin{align}
    \mathcal{H}&=\frac{1}{2}m\eta v^2-\frac{Gm^2\eta}{r}\mathcal{Z}\notag\\
    &+\frac{1}{c^2}\bigg\{\frac{3}{8}m\eta(1-3\eta)v^4+\frac{1}{2}\frac{G m^2\eta}{r}\bigg[\frac{G m}{r}+\dot{r}^2\eta+(3+\eta)v^2-\mathcal{Q}\bigg(\eta\dot{r}^2+\eta v^2+\frac{G m}{r}(2-\zeta_0)\bigg)\bigg]\bigg\}+\mathcal{O}(c^{-4}),\\
    \mathcal{J}^i&=\bigg\{1+\frac{1}{c^2}\bigg[3\frac{G m}{r}+\eta\frac{G m}{r}\mathcal{Z}+\frac{1}{2}(1-3\eta)v^2\bigg]\bigg\}m\eta\omega r^2 \hat{k}^i+\mathcal{O}(c^{-4}).\label{AngularMomentum}
\end{align}
\end{widetext}
Here we use some identities about $\chi_1$ and $\chi_2$ which are
\begin{gather}
    \chi_1+\chi_2=1,\quad \chi_1^2+\chi_2^2=1-2\eta,\notag\\
    \chi_1^3+\chi_2^3=1-3\eta,
\end{gather}
where the unit vector $\hat{k}^i$ is the normal to the orbital plane. We use the relationship $\epsilon_{ijk}x^jv^k=r^2\omega\hat{k}^i$, and $\omega$ is the orbital angular frequency. We obtain the acceleration
\begin{align}
\label{aCoM}
    \bm{a}=&-\frac{Gm}{r^2}\bigg\{\left[1- \mathcal{Q}+\frac{1}{c^2}\Big(\mathcal{A}_g- \mathcal{Q}\mathcal{A}_{EM}\Big)\right]\bm{n}\notag\\
    &+\frac{1}{c^2}\Big[\mathcal{B}_g-\mathcal{Q}\mathcal{B}_{EM}\Big]\bm{v}\bigg\}+\mathcal{O}(c^{-4}),
\end{align}
we have $\dot{r}=\bm{v}\cdot\bm{n}$, and the coefficients are
\begin{align}
    \mathcal{A}_g&=-\frac{3}{2}\eta\dot{r}^2+v^2+3\eta v^2-\frac{Gm}{r}(4+2\eta+{\zeta}_0),\\
    \mathcal{A}_{EM}&=-\frac{3}{2}\eta \dot{r}^2-\frac{1}{2}v^2-\frac{Gm}{r}(5+4\eta-2\eta\mathcal{Q}),\\[6pt]
    \mathcal{B}_g&=-4\dot{r}+2\eta\dot{r},\\[10pt]
    \mathcal{B}_{EM}&=\dot{r}-2\eta\dot{r},
\end{align}
where the dimensionless coefficients $\zeta_0$ is given by
\begin{align}
    {\zeta}_0\equiv\frac{  q_1^2}{ Gm_1m}+\frac{ q_2^2}{Gm_2m}.
\end{align}

In the center-of-mass frame, the conservative acceleration is computed through 1PN order. The radiative effects enter the subsequent flux-balance analysis starting at 1.5PN order. It is difficult to obtain the 1.5PN, 2.5PN, and 3.5PN acceleration in the harmonic gauge; the three-dimensional radiation-reaction acceleration is gauge dependent, such as the Burke-Thorne gauge and the Damour-Deruelle gauge \cite{Maggiore:2007ulw,Gravity_PoissonWill,PhysRevD.47.4392}, which construct odd PN orders of metric by the multipole moments of mass and charge. In the next section, we calculate the radiation-driven evolution of the orbital angular frequency $\dot{\omega}$ as a gauge-invariant  quantity. We obtain $\dot{\omega}$ from angular-momentum flux balance.

\section{Applications in circular orbits}\label{Sec4}

\subsection{Kepler's third law}
We focus on quasi-circular binaries and use angular-momentum flux balance to determine the frequency evolution. We rewrite the velocity in   {the orthonormal polar basis} $\bm{v}=\dot{r}\bm{n}+r\omega\bm{\tau}$, $\bm{\tau}$ is the unit tangent vector in the orbit plane, $\bm{n}\cdot\bm{\tau}=0$, and $\omega$ is the angular frequency of binaries. The basis vectors obey
\begin{equation}\label{orthogonalframe}
    \dot{\bm{n}}=\omega\bm{\tau},\qquad\dot{\bm{\tau}}=-\omega\bm{n},
\end{equation}
and the acceleration is
\begin{gather}
    \bm{a}=\bm{\dot{v}}=(\ddot{r}-\omega^2r)\bm{n}+(\dot{\omega}r+2\omega \dot{r})\bm{\tau},\\
    \bm{a}\cdot\bm{n}=\ddot{r}-\omega^2r,\\
    \bm{a}\cdot\bm{v}=\ddot{r}\dot{r}+\omega^2r\dot{r}+\omega\dot{\omega}r^2,
\end{gather}
Projecting the acceleration \eqref{aCoM} along the radial direction and setting $\dot{r}=0$ yields
\begin{align}
    \omega^2 r=\frac{Gm}{r^2}\bigg[1-\mathcal{Q}+\frac{1}{c^2}\Big({\mathcal{A}}_g-\mathcal{Q}\mathcal{A}_{EM}\Big)\bigg]
\end{align}
We obtain Kepler’s third law with the EM correction at 1PN order as
\begin{align}
    \omega^2&=\frac{Gm}{r^3}\bigg\{1-\mathcal{Q}+\frac{1}{c^2}\frac{Gm}{r}\bigg[\eta-3-\frac{1}{2}(4\eta+1)\mathcal{Q}^2\notag\\
    &+\bigg(\frac{9}{2}+\eta\bigg)\mathcal{Q}-\zeta_0\bigg]\bigg\}+\mathcal{O}(c^{-4}).\label{omega}\\
    r&=\Big(\frac{Gm}{\omega^2}\mathcal{Z}\Big)^{1/3}\bigg\{1+\frac{x}{\mathcal{Z}^{1/3}}\frac{1}{3}\bigg[\eta-3-\frac{1}{2}(4\eta+1)\mathcal{Q}^2\notag\\
    &+\bigg(\frac{9}{2}+\eta\bigg)\mathcal{Q}-\zeta_0\bigg]\bigg\}+\mathcal{O}(c^{-4}).\label{ro}
\end{align}
Taking the derivative of both sides of this equation with respect to time, we obtain 
\begin{equation}
    \frac{\dot{r}}{r}=-\frac{2}{3}\frac{\dot{\omega}}{\omega}+\mathcal{O}(c^{-4}).
\end{equation}
where the PN parameter is
\begin{equation}\label{Omega}
    x=\frac{(G m \omega)^{2/3}}{c^2},
\end{equation}
The energy $\mathcal{E}_c$ and the angular momentum $\mathcal{J}_c\equiv|\mathcal{J}_c^i|=|\mu\epsilon_{ijk} x_j v_k|$ in the circular orbits are 
\begin{align}
    \mathcal{E}_c=&-\frac{\mu c^2}{2}\frac{x}{\mathcal{Z}^{1/3}}\bigg\{\mathcal{Z} + \frac{x}{\mathcal{Z}^{1/3}} \bigg[-\frac{3}{4}-\frac{\eta }{12}\notag\\
    &+\mathcal{Q} \left(-\frac{1}{2}+\frac{\zeta _0}{3}-\frac{7 \eta }{2}\right)+\mathcal{Q}^2 \left(\frac{7}{12}+\frac{17}{4}-\frac{\zeta_0}{3}\right)\notag\\
    &+\mathcal{Q}^3\left(-\frac{2 \eta }{3}-\frac{1}{6}\right)\bigg]\bigg\}+\mathcal{O}(c^{-4}),\\
    \mathcal{J}_c=&\frac{G m\mu}{c\sqrt{x}}\mathcal{Z}^{2/3}  \bigg\{1+\frac{x}{\mathcal{Z}^{1/3}}\bigg[\frac{3}{2}+\frac{\eta }{6}+\mathcal{Q} \left(\frac{5}{2}-\frac{2 \zeta _0}{3}+\frac{7\eta }{6}\right)\notag\\
    &+\left(-\frac{4 \eta }{3}-\frac{1}{3}\right) \mathcal{Q}^2\bigg]\bigg\}+\mathcal{O}(c^{-4}).
\end{align}
  {Each PN order $(c^{-2})$ corresponds to one power of $x$  in the circular orbit.}

\subsection{The gravitational and electromagnetic radiation in circular orbits}

In the presence of EM and GW radiation, because the EM contribution enters at the 1.5PN order and GW contribution enters at the 2.5PN order, we use the flux-balance equations \footnote{The flux–balance equations hold in an orbit-averaged sense. See Sec. 15 in \cite{damour1983gravitational}, Eq. (184) in \cite{Damour1987}, and Eqs. (4.6)-(4.7) in \cite{deAndrade2001}. } \cite{PhysRevD.102.103520,PhysRevD.42.1123, Liu2021,Blanchet_2019}
\begin{align}
    \frac{\mathrm{d}\mathcal{H}}{\mathrm{d}t}&=-\mathcal{F}_{GW}-\mathcal{F}_{EM},\quad
    \frac{\mathrm{d}\mathcal{J}^i}{\mathrm{d}t}=-\mathcal{G}^i_{GW}-\mathcal{G}^i_{EM},\notag\\
    \frac{\mathrm{d}\mathcal{P}^i}{\mathrm{d}t}&=0,\quad \frac{\mathrm{d}\mathcal{N}^i}{\mathrm{d}t}=\mathcal{P}^i-\Psi^i_{EM},
\end{align}
where $\Psi_{EM}^i$ comes from the EM terms in Eq. \eqref{PG}, which can be seen in \cite{PhysRevD.108.024020} as well. 

For quasi-circular orbits, the evolution of the orbital angular frequency can be derived equivalently from either the angular-momentum flux balance equation or the energy flux balance equation with identical results. From the time derivative of the angular momentum of 1PN in Eq. \eqref{AngularMomentum}, we obtain
\begin{align}
\frac{\mathrm{d}\mathcal{J}_c}{\mathrm{d}t}&=-\frac{1}{3}\mu \Big(\frac{G m}{\omega^2}\mathcal{Z}\Big)^{2/3}\dot{\omega}\bigg\{1+\frac{1}{2}\mathcal{Z}x\bigg[3+\frac{1}{3} \eta \notag\\
    &-\frac{2}{3}(1+4\eta)\mathcal{Q}^2+ \frac{1}{3}\left(15+7\eta\right)\mathcal{Q}-\frac{4}{3}\zeta_0\bigg]\bigg\}+\mathcal{O}(c^{-4}),
\end{align}
where $\mu=m\eta$ is the reduced mass. We have used $\dot{\omega}$ to replace $\dot{r}$ and neglect terms of order $\mathcal{O}(\dot{\omega}^2)\thicksim \mathcal{O}(c^{-6})$. 

 Using the leading-order matching between radiative multipoles in the multipolar post-Minkowskian expansion and source multipoles in the 1PN expansion in Appendix \ref{App.B}, we express the EM and GW radiation fluxes in terms of  multipole moments as follows

\begin{widetext}

\begin{align}
\mathcal{F}_{GW}=&\frac{G}{c^5}\frac{1}{5}\big\langle \hat{\mathrm{I}}^{(3)}_{ij}\hat{\mathrm{I}}^{(3)}_{ij}\big\rangle+\frac{G}{c^7}\bigg[\frac{16}{45}\big\langle \hat{\mathrm{J}}^{(3)}_{ij}\hat{\mathrm{J}}^{(3)}_{ij}\big\rangle+\frac{1}{189}\big\langle \hat{\mathrm{I}}^{(4)}_{ijk}\hat{\mathrm{I}}^{(4)}_{ijk}\big\rangle\bigg]+\mathcal{O}(c^{-9}),\\
\mathcal{G}^i_{GW}=&\frac{G}{c^5}\frac{2}{5}\big\langle \hat{\mathrm{I}}^{(2)}_{jl}\hat{\mathrm{I}}^{(3)}_{kl}\big\rangle\epsilon_{ijk}+\frac{G}{c^7}\bigg[\frac{32}{45}\big\langle \hat{\mathrm{J}}^{(2)}_{jl}\hat{\mathrm{J}}^{(3)}_{kl}\big\rangle+\frac{1}{63}\big\langle \hat{\mathrm{I}}^{(3)}_{jlm}\hat{\mathrm{I}}^{(4)}_{klm}\big\rangle\bigg]\epsilon_{ijk}+\mathcal{O}(c^{-9}),\\
\mathcal{F}_{EM}=&\frac{G}{c^3}\frac{2}{3}\big\langle {\mathrm{E}}^{(2)}_{i}{\mathrm{E}}^{(2)}_{i}\big\rangle
+\frac{G}{c^5}\bigg[\frac{1}{6}\big\langle \hat{\mathrm{B}}^{(2)}_{i}\hat{\mathrm{B}}^{(2)}_{i}\big\rangle+\frac{1}{20}\big\langle \hat{\mathrm{E}}^{(3)}_{ij}\hat{\mathrm{E}}^{(3)}_{ij}\big\rangle\bigg]
+\frac{G}{c^7}\bigg[\frac{1}{45}\big\langle \hat{\mathrm{B}}^{(3)}_{ij}\hat{\mathrm{B}}^{(3)}_{ij}\big\rangle+\frac{2}{945}\big\langle \hat{\mathrm{E}}^{(4)}_{ijk}\hat{\mathrm{E}}^{(4)}_{ijk}\big\rangle\bigg]\notag\\
&+\mathcal{O}(c^{-9}),\\
\mathcal{G}^i_{EM}=&\frac{G}{c^3}\frac{2}{3}\big\langle {\mathrm{E}}^{(1)}_{j}{\mathrm{E}}^{(2)}_{k}\big\rangle\epsilon_{ijk}
+\frac{G}{c^5}\bigg[\frac{1}{6}\big\langle \hat{\mathrm{B}}^{(1)}_{j}\hat{\mathrm{B}}^{(2)}_{k}\big\rangle+\frac{1}{10}\big\langle \hat{\mathrm{E}}^{(2)}_{jl}\hat{\mathrm{E}}^{(3)}_{kl}\big\rangle\bigg]\epsilon_{ijk}+\frac{G}{c^7}\bigg[\frac{2}{45}\big\langle \hat{\mathrm{B}}^{(2)}_{jl}\hat{\mathrm{B}}^{(3)}_{kl}\big\rangle+\frac{2}{315}\big\langle \hat{\mathrm{E}}^{(3)}_{jlm}\hat{\mathrm{E}}^{(4)}_{klm}\big\rangle\bigg]\epsilon_{ijk}\notag\\
&+\mathcal{O}(c^{-9}),
\end{align}

For the radiation parts on the right side of the balance equation, using the multipole expansion derived in Appendix \ref{App.B} we obtain the multipole moments at the center of mass frame as \cite{PhysRevD.43.3259,1989MNRAS.239..845B,7mqv-rhc6}

\begin{align}\label{Multipoles}
    \mathrm{E}_{i}=&\mu\xi_1x_i+\frac{\mu}{c^2}\bigg\{\bigg[-\frac{1}{2}\xi_0\frac{Gm}{r}\eta\mathcal{Z}-\xi_2\frac{Gm}{r}-\frac{1}{10}\xi_3\frac{G m}{r}\mathcal{Z}+\frac{1}{2}\xi_3 v^2\bigg]x_i+\frac{3}{10}\xi_3 r\dot{r}v_i\bigg\}+\mathcal{O}(c^{-4}),\\
    \mathrm{B}_i=&\mu\epsilon_{ijk}x_{\langle j}v_{k\rangle}\zeta_2+\frac{\mu}{c^2}\epsilon_{ijk}x_{\langle j}v_{k\rangle}\bigg\{\zeta_4\bigg(2v^2-2\frac{Gm}{r}\mathcal{Z}\bigg)-\zeta_3\frac{Gm}{r}\bigg\}+\mathcal{O}(c^{-4}),\\
    \hat{\mathrm{E}}_{ij}=&\mu\zeta_2 x_{\langle i}x_{j\rangle}+\frac{\mu}{c^2}\bigg\{\bigg[-\zeta_3\frac{G m}{r}+\zeta_4\Big(-\frac{3}{7}\frac{Gm}{r}\mathcal{Z}+\frac{8}{21}v^2\Big)\bigg]x_{\langle i}x_{j\rangle}+\frac{6}{7}\zeta_4r\dot{r}x_{\langle i}v_{j\rangle}+\frac{1}{21}\zeta_4r^2v_{\langle i}v_{j\rangle}\bigg\}+\mathcal{O}(c^{-4}),\\
    \hat{\mathrm{I}}_{ij}=&\mu x_{\langle i}x_{j\rangle}+\frac{\mu}{c^2}\bigg\{\bigg[\frac{29}{42}(1-3\eta)v^2-\frac{1}{7}(5-8\eta)\frac{G m}{r}\mathcal{Z}\bigg]x_{\langle i}x_{j\rangle}-\frac{4}{7}(1-3\eta)r\dot{r}x_{\langle i}v_{j\rangle}+\frac{11}{21}(1-3\eta)r^2v_{\langle i}v_{j\rangle}\bigg\}+\mathcal{O}(c^{-4}).
\end{align}
\end{widetext}
The symbol ``$\hat{\quad}$'' denotes the symmetric trace-free (STF) tensor, and 
\begin{equation}
      x_{\langle i}x_{j\rangle}=x_{( i}x_{j)}-\frac{1}{3}\delta_{ij}x_kx^k.
\end{equation}
The electric dipole current, quadrupole currents, and octupole at Newtonian order are
\begin{align}
    \hat{\mathrm{B}}_{ij}=&\mu\xi_3\epsilon_{kl\langle i}x_{j\rangle}x_k v_{l},\\
    \hat{\mathrm{E}}_{ijk}=&\mu\xi_3x_{\langle i}x^j x_{k\rangle}.
\end{align}
The mass quadrupole currents and octupoles of Newtonian order are
\begin{align}
    \hat{\mathrm{J}}_{ij}=&\mu\sqrt{1-4\eta}\epsilon_{kl\langle i}x_{j\rangle}x_k v_{l},\\
    \hat{\mathrm{I}}_{ijk}=&\mu\sqrt{1-4\eta}x_{\langle i}x_j x_{k\rangle}.
\end{align}
Because $a^i\propto x^i$ at the Newtonian order and $\epsilon_{ijk} x^jx^k=0$ and $\epsilon_{ijk} v^jv^k=0$, we obtain that the electric current dipole $\dot{\mathrm{B}}_i=\mathcal{O}(c^{-5})$ or ${\mathrm{B}}_i$ is constant in the circular orbits.

We denote charge-to-mass ratios as $\kappa_1= q_1/m_1$ and $\kappa_2= q_2/m_2$, which have a relationship to $\mathcal{Q}$ and $\mathcal{\zeta}_0$
\begin{equation}
\mathcal{Q}=\kappa_1\kappa_2\quad \text{and} \quad\zeta_0=\chi_1\kappa_1^2+ \chi_2\kappa_2^2,
\end{equation}
where $|\kappa_1|,|\kappa_2|<1$. $\kappa<0$ means a black hole with negative charge, while $\kappa>0$ is positive charge. We set dimensionless coefficients combined with charge $q_1, q_2$ and dimensionless mass ratio $\chi_1, \chi_2$
\begin{gather}
\xi_0\equiv(\kappa_1\chi_1+\kappa_2\chi_2)(\chi_1-\chi_2),\notag\\
\xi_1\equiv\kappa_1-\kappa_2,\quad\xi_2\equiv\kappa_1\chi_2-\kappa_2\chi_1,\notag\\
\xi_3\equiv\kappa_1\chi_2^2-\kappa_2\chi_1^2,\notag\\
\zeta_2\equiv\kappa_1\chi_2+\kappa_2\chi_1,\quad
\zeta_3\equiv\kappa_1\chi_2^2+\kappa_2\chi_1^2,\notag\\
\zeta_4\equiv\kappa_1\chi_2^3+\kappa_2\chi_1^3,
\end{gather}
where $\xi_0$ comes from the center of mass frame in Eq. \eqref{xCoM}. In particular, if $m_1=m_2$ and $\eta=1/4$, the current quadrupole and the mass octupole $\mathrm{J}^{ij}=\mathrm{I}^{ijk}=0$. Based on the equal mass condition, if $q_1=q_2$ the electric dipole $\mathrm{E}^i=0$, and if $q_1=-q_2$ the electric quadrupole $\mathrm{E}^{ij}=0$.

\begin{widetext}
Substituting these mass and electric multipoles into the energy and angular momentum radiation, we obtain

\begin{align}
    \mathcal{F}_{GW}&=\frac{32}{5} \frac{c^5}{G}\eta^2  \mathcal{Z}^{4/3}x^{5}\bigg\{1+\frac{x}{\mathcal{Z}^{4/3}} \bigg[-8-4 \eta -2\zeta_0-4 \eta  \mathcal{Q}^2+ ( 10+8 \eta )\mathcal{Q}\notag\\
    &\qquad\qquad+\frac{2}{3}\bigg(6+\zeta_0+8 \eta +\frac{1}{2}(4\eta+1)\mathcal{Q}^2-(3+\eta)\mathcal{Q}\bigg)\mathcal{Z}+\left(\frac{97}{336}-\frac{17 \eta }{4}\right) \mathcal{Z}^{2}\bigg]\bigg\}+\mathcal{O}(c^{-9}),\\
    \mathcal{F}_{EM}&=\frac{2}{3} \frac{c^5}{G}\eta^2 
    \mathcal{Z}^{2/3}x^4\bigg\{\xi_1^2+\frac{x}{\mathcal{Z}^{4/3}} \bigg[\bigg( \left(10  +8 \eta \right)\mathcal{Q}-4\eta -8-2\zeta_0 -4 \eta  \mathcal{Q}^2\bigg)\xi_1^2+\Big(\frac{12}{5}\zeta_2^2-\eta\xi_0\xi_1+\frac{4}{5}\xi_1\xi_3\Big)\mathcal{Z}^2\notag\\
    &\qquad\qquad+\frac{2}{3}\bigg(\Big(9+7 \eta +2\zeta_0+(1+4\eta)\mathcal{Q}^2+\left(-\frac{15}{2} -2 \eta  \right)\mathcal{Q}\Big) \xi_1^2-\frac{1}{3} \xi_1\xi_2\bigg)\mathcal{Z}\bigg]\notag\\
    &+\frac{24}{5}\frac{x^2}{\mathcal{Z}^{8/3}}\bigg[2\bigg(-2- \eta -  \eta \mathcal{Q}^2+ \zeta_0+ ( 2 \eta +1)\mathcal{Q}\bigg)\zeta_2^2+\bigg(-\frac{2}{21}\zeta_2\zeta_4-\frac{1219}{1120}(4 \eta -1)\xi_3^2\bigg)\mathcal{Z}^2\notag\\
    &\qquad\qquad+\bigg(\Big(2+\frac{8}{3}\eta+\frac{1}{3}\zeta_0+\frac{1}{6} (1+4\eta)\mathcal{Q}^2+\frac{1}{3}\left(3+ \eta\right)\mathcal{Q}\Big)\zeta_2^2- \zeta_2\zeta_3\bigg)\mathcal{Z}\bigg]\bigg\}+\mathcal{O}(c^{-9}),
\end{align}
and the angular momentum radiation are
 \begin{align}
    \mathcal{G}^i_{GW}&=\frac{32}{5}\frac{c^3}{G}\eta^2   \mathcal{Z}^{2/3}x^{4}\frac{\mathcal{J}_c^i}{\mu}\bigg\{1+\frac{x}{\mathcal{Z}^{4/3}} \bigg[-6-3 \eta-\frac{3}{2}\zeta_0 -3 \eta  \mathcal{Q}^2+\frac{3}{2} (5+4 \eta)\mathcal{Q}\notag\\
    &\qquad\qquad+\bigg(4+\frac{11}{3} \eta+\frac{5}{6}\zeta_0 +\frac{5}{12}(1+4\eta)\mathcal{Q}^2-\left(3+\frac{5}{6}\eta \right)\mathcal{Q} \bigg)\mathcal{Z}+\left(\frac{97}{336}-\frac{17 }{4}\eta\right) \mathcal{Z}^{2}\bigg]\bigg\}+\mathcal{O}(c^{-9}),\\
    \mathcal{G}^i_{EM}&=\frac{2}{3}\frac{c^3}{ G}   \eta^2 x^{3}\frac{\mathcal{J}^i_c}{\mu}\bigg\{\xi_1^2+\frac{x}{\mathcal{Z}^{4/3}} \bigg[\bigg(-2\eta -4 -\zeta_0- 2\eta  \mathcal{Q}^2+ \left(5 +4 \eta \right)\mathcal{Q}\bigg)\xi_1^2+\Big(-\xi_0\xi_1+\frac{4}{5}\xi_1\xi_3+\frac{12}{5}\zeta_2^2\Big)\mathcal{Z}^2\notag\\
    &\qquad\qquad+\bigg(\Big(4+2\eta +\zeta_0+\frac{1}{2}(1+4\eta)\mathcal{Q}^2-\left( 4+\eta  \right)\mathcal{Q}\Big) \xi_1^2-2 \xi_1\xi_2\bigg)\mathcal{Z}\bigg]\notag\\
    &+\frac{24}{5}\frac{x^2}{\mathcal{Z}^{8/3}}\bigg[\frac{3}{2}\bigg(-2- \eta+\zeta_0 -  \eta \mathcal{Q}^2+  ( 2 \eta +1)\mathcal{Q}\bigg)\zeta_2^2+\bigg(-\frac{2}{21}\zeta_2\zeta_4+\frac{1219}{1120}(4 \eta -1)\xi_3^2\bigg)\mathcal{Z}^2\notag\\
    &\qquad\qquad+\bigg(\Big(2+\frac{11}{6}\eta+\frac{5}{12}\zeta_0+\frac{5}{24}(1+4\eta) \mathcal{Q}^2-\Big(\frac{3}{2}+ \frac{5}{12}\eta\Big)\mathcal{Q}\Big)\zeta_2^2- \zeta_2\zeta_3\bigg)\mathcal{Z}\bigg]\bigg\}+\mathcal{O}(c^{-9}).
\end{align}
 {Note that the conservative dynamics and the source multipoles used in this work, including Kepler's law, the current, orbital binding energy, and the electric dipole moment, are computed through the next-to-leading order (NLO). Because the leading electric-dipole radiation appears at $-1$ PN order, the electric-dipole contribution derived here is accurate up to 0 PN order. A complete 1 PN-accurate treatment would require the next-to-next-to-leading order (NNLO) electric-dipole radiation and the 2PN conservative dynamics, which are beyond the scope of the present calculation.} 
Substituting the angular momentum radiation $\mathcal{G}^i=\mathcal{G}^i_{GW}+\mathcal{G}^i_{EM}$ in the balance equation, we obtain the variation of the orbital angular frequency
\begin{align}\label{Domega}
    \dot{\omega}&=\frac{G \mu}{c^3}\omega^{3}\bigg\{2\xi_1^2+\frac{x}{\mathcal{Z}^{4/3}}\bigg(\frac{96}{5}+\sum_{a=0}^3\mathcal{C}_a\mathcal{Q}^a \bigg)+\frac{x^2}{\mathcal{Z}^{8/3}}\bigg( -\frac{264 }{5}\eta-\frac{1486}{35}+\sum_{a=0}^6\mathcal{D}_a\mathcal{Q}^a\bigg)\bigg\}+\mathcal{O}(c^{-9}),
\end{align}
where $\mathcal{Q}^a$ denotes the $a$-th power of $\mathcal{Q}$, $\mathcal{C}_a$ and $\mathcal{D}_a$ are
\begin{align}
    \mathcal{C}_0=&\xi _1^2 \left(\frac{5 \zeta _0}{3}+\frac{5 \eta }{3}-1\right)+\frac{24 \zeta _2^2}{5}-2 \xi _0 \xi _1-8 \xi _1 \xi _2,\\
     \mathcal{C}_1=&\xi _1^2 \left(-\zeta _0+4 \eta +6\right)-\frac{48}{5} \zeta _2^2+2 \xi _0 \xi _1+\xi _1 \left(8 \xi _2-2 \xi _3\right)-\frac{192}{5},\\
     \mathcal{C}_2=& \xi _1^2\left(\frac{4 \zeta _0}{3}-11 \eta -\frac{28}{3}\right)+\frac{24 \zeta _2^2}{5}+2 \xi _1 \xi _3+\frac{96}{5},\\
     \mathcal{C}_3=&\xi _1^2\left(\frac{16 \eta }{3}+\frac{1}{3}\right) , \\
    \mathcal{D}_0=&\xi _1^2 \left(\zeta _0 \left(\frac{11 \eta }{6}-\frac{13}{6}\right)-\frac{16 \zeta _0^2}{9}+\frac{\eta ^2}{2}+\eta +\frac{1}{2}\right)+\xi _0 \xi _1 \left(\frac{4 \zeta _0}{3}-\frac{5 \eta }{3}+1\right)+\xi _1 \xi _2 \left(\frac{16 \zeta _0}{3}-\frac{20 \eta }{3}+4\right) \notag\\
    &+\zeta _2^2\left(\frac{36 \eta }{5}-12\right)+\left(-\frac{96 \zeta _3}{5}-\frac{16 \zeta _4}{7}\right) \zeta _2+\zeta _0 \left(6 \zeta _2^2+\frac{16}{5}\right)+\left(\frac{7314 \eta }{175}-\frac{3657}{350}\right) \xi _3^2,\\
    \mathcal{D}_1=&\xi _1^2 \left(\zeta _0 \left(\frac{103}{6}-\frac{7 \eta }{6}\right)+\frac{10 \zeta _0^2}{9}+\frac{11 \eta ^2}{3}+3 \eta +10\right)+\xi _0 \xi _1 \left(-\frac{4 \zeta _0}{3}-\frac{\eta }{3}-13\right)+\xi _1\xi _2 \left(-\frac{16 \zeta _0}{3}-\frac{4 \eta }{3}-52\right)\notag\\
    &+\xi _1\xi _3 \left(\frac{4 \zeta _0}{3}-\frac{5 \eta }{3}+1\right)+\zeta _2^2 \left(\frac{228}{5}-\frac{12 \eta }{5}\right)+\left(\frac{288 \zeta _3}{5}+\frac{64 \zeta _4}{7}\right) \zeta _2+\zeta _0 \left(\frac{176}{5}-14 \zeta _2^2\right)\notag\\
    &+\left(\frac{7314}{175}-\frac{29256 \eta }{175}\right) \xi _3^2+\frac{1584 \eta }{5}+\frac{6952}{35},\\
    \mathcal{D}_2=&\xi _1^2 \left(\zeta _0 \left(-\frac{17 \eta }{2}-\frac{143}{6}\right)+\frac{4 \zeta _0^2}{3}-\frac{14 \eta ^2}{3}+\frac{15 \eta }{2}-22\right)+\xi _0 \xi _1 \left(-\frac{4 \zeta _0}{3}+11 \eta +\frac{73}{3}\right)+\xi _1\xi _2 \left(-\frac{16 \zeta _0}{3}+44 \eta +\frac{292}{3}\right)\notag\\
    &+\xi _1\xi _3 \left(-\frac{4 \zeta _0}{3}-\frac{\eta }{3}-13\right)+\zeta _2^2 \left(-\frac{244 \eta }{5}-\frac{402}{5}\right)+\left(-\frac{288 \zeta _3}{5}-\frac{96 \zeta _4}{7}\right) \zeta _2+\left(\frac{66 \zeta _2^2}{5}+\frac{96}{5}\right) \zeta _0\notag\\
    &+\left(\frac{43884 \eta }{175}-\frac{10971}{175}\right) \xi _3^2-\frac{3264 \eta }{5}-\frac{2220}{7},\\
    \mathcal{D}_3=&\xi _1^2 \left(\zeta _0 \left(\frac{37 \eta }{2}+\frac{39}{2}\right)-\frac{14 \zeta _0^2}{9}-21 \eta ^2-\frac{343 \eta }{6}-\frac{13}{2}\right)+\xi _0 \xi _1 \left(\frac{4 \zeta _0}{3}-\frac{43 \eta }{3}-\frac{41}{3}\right)+\xi _1 \xi _2 \left(\frac{16 \zeta _0}{3}-\frac{172 \eta }{3}-\frac{164}{3}\right)\notag\\
    &+\xi _1\xi _3 \left(-\frac{4 \zeta _0}{3}+11 \eta +\frac{73}{3}\right)+\zeta _2^2 \left(\frac{444 \eta }{5}+\frac{366}{5}\right)+\left(\frac{96 \zeta _3}{5}+\frac{64 \zeta _4}{7}\right) \zeta _2+\left(-\frac{1}{5} 42 \zeta _2^2-\frac{208}{5}\right) \zeta _0\notag\\
    &+\left(\frac{7314}{175}-\frac{29256 \eta }{175}\right) \xi _3^2+\frac{2928 \eta }{5}+\frac{7456}{35},\\
    \mathcal{D}_4=& \xi _1^2 \left(\zeta _0 \left(-16 \eta -\frac{34}{3}\right)+\frac{8 \zeta _0^2}{9}+\frac{281 \eta ^2}{6}+\frac{457 \eta }{6}+19\right)+\xi _0 \xi _1\left(\frac{16 \eta }{3}+\frac{4}{3}\right) +\xi _1\xi _2\left(\frac{64 \eta }{3}+\frac{16}{3}\right)\notag\\
    &+\xi _1 \xi _3 \left(\frac{4 \zeta _0}{3}-\frac{43 \eta }{3}-\frac{41}{3}\right)+\zeta _2^2 \left(-\frac{288 \eta }{5}-\frac{138}{5}\right)-\frac{16 \zeta _4 \zeta _2}{7}+\left(\frac{16 \zeta _2^2}{5}+\frac{64}{5}\right) \zeta _0\notag\\
    &+\left(\frac{7314 \eta }{175}-\frac{3657}{350}\right) \xi _3^2-216 \eta -\frac{398}{7},\\
     \mathcal{D}_5=& \xi _1^2 \left(\zeta _0 \left(\frac{16 \eta }{3}+\frac{2}{3}\right)-\frac{100 \eta ^2}{3}-\frac{191 \eta }{6}-\frac{5}{6}\right)+\zeta _2^2 \left(\frac{64 \eta }{5}+\frac{6}{5}\right)+\left(\frac{16 \eta }{3}+\frac{4}{3}\right) \xi _1 \xi _3+\frac{96 \eta }{5}+\frac{24}{5},\\
    \mathcal{D}_6=&\xi _1^2\left(8 \eta ^2+\frac{4 \eta }{3}-\frac{1}{6}\right),
\end{align}
\end{widetext}
The leading order contribution to $\dot{\omega}$, proportional to $\xi_1^2$, arises from electric-dipole radiation and enters at -1 PN order relative to the leading order GW radiation. The term proportional to $x^2$ represents part of the 1 PN order contribution. Upon completion of the 1 PN order calculation, the coefficients $\mathcal{D}_a$ will be modified accordingly.  Placidi et al. \cite{placidi2025chargedblackholebinaryevolution} and Tang \cite{tang2025inspiralingbinarychargedblack} derived the PN Lagrangian and analyzed the leading order of EM radiation. Here we obtain the radiation-driven evolution including the EM and GW contributions at the stated PN accuracy. As expected, when setting $\mathcal{Z}=1, \mathcal{Q}=0$, and all other electric coupling coefficients $\xi_i=\zeta_i=0$, the electric effects are eliminated and $\dot{\omega}$ reduces to the standard PN results. If $q_1\neq0$ and $q_2=0$, the system reduces to a binary composed of a charged BH and a neutral BH. In this case, there is EM radiation due to the motion of $q_1$ without the Coulomb interaction. 

\subsection{The innermost stable circular orbit}
In this section, we discuss the stability of a binary system in a circular orbit. The derivation of the innermost stable circular orbit (ISCO) for a test particle in the Reissner–Nordström spacetime is a standard exercise in general relativity \cite{PhysRevD.88.024042,PhysRevD.83.104052,PhysRevD.77.103005}. By extracting the effective potential $V_{eff}(r)$ and finding the circular orbit condition $\partial_r V_{eff}(r)=0$ and the stable condition $\partial_r^2 V_{eff}(r)=0$, we derive an equation for the ISCO radius  $r_{ISCO}$ following the work of Chandrasekhar \cite{Chandrasekhar}. An analytic expression for the ISCO of a neutral test particle in the RN spacetime is cumbersome and does not directly describe comparable-mass binaries. In the test-mass limit $\eta\rightarrow 0$, the PN ISCO should reduce to the test-particle result in the RN metric, which we have discussed in Appendix \ref{App.C}. We now determine the binary ISCO within the PN framework \cite{Blanchet2024,LucBlanchet_2003,VanessaCdeAndrade_2001}. 
The conjugate momentum is
\begin{equation}
    p^i=\frac{1}{\mu}\frac{\partial L}{\partial v^i}, \quad p_R\equiv \bm{n}\cdot \bm{p}.
\end{equation}
In the spherical coordinate system ($r$,$\theta$,$\phi$), we obtain
\begin{equation}
    p^2=p_r^2+\frac{p_\phi^2}{r^2},\quad p_\phi=r^2\bm{p}\cdot\bm{\tau},
\end{equation}
where $p_{\phi}$   {corresponds to the angular momentum per unit reduced mass. }
Performing the Legendre transform, we obtain the Hamiltonian as a function of $\bm{p}$
    \begin{align}
    \frac{\mathcal{H}}{\mu}&=-\frac{G m}{r}\mathcal{Z}+\frac{1}{2}p^2\notag\\
    &+\frac{1}{c^2}\bigg\{\frac{1}{8}(3\eta-1)p^4+\frac{1}{2}\frac{G m}{r}\bigg[\frac{G m}{r}\Big(1-2\mathcal{Q}+\zeta_0\Big)\notag\\
    &-\eta p_r^2-(3+\eta)p^2+\mathcal{Q}(p_r^2+p^2)\bigg]\bigg\}+\mathcal{O}(c^{-4}).\label{Hamitonian}
\end{align}

This Hamiltonian is consistent with \cite{PhysRevD.57.7274,PhysRevD.62.084011}  by setting $\mathcal{Z}=1$ and $\mathcal{Q}=0$. We denote $\bar{\mathcal{H}}\equiv \mathcal{H}/\mu$. 
The Hamiltonian $\bar{\mathcal{H}}[r,p_r,p_\phi]$ is a function of $r,p_r$ and $p_\phi$. For a circular orbit with radius $r=r_0$, the radial force balance condition is
\begin{equation}
    \frac{\partial \bar{\mathcal{H}}}{\partial r}\big[r_0,0,p_\phi^0\big]=0,
\end{equation}
which determines $p_\phi^0$.   {$\partial\bar{\mathcal{H}}/\partial p_r=0$ is satisfied at $p_r=0$.} The angular frequency is determined by
\begin{equation}
     \omega_0=\frac{\mathrm{d}\phi}{\mathrm{d}t}\Big|_{r0}=\frac{\partial \bar{\mathcal{H}}}{\partial p_\phi}\big[r_0,0,p_\phi^0\big],
\end{equation}
The resulting expression for $\omega_0$ agrees with Eq.~\eqref{omega}. Considering a perturbation of the circular orbit,
\begin{align}
    p_r&=\delta p_r,\qquad\; p_\phi=p_\phi^0+\delta p_\phi,\notag\\
    r&=r_0+\delta r,\quad \omega=\omega_0+\delta \omega.
\end{align}
In linear order of the perturbation, we have
\begin{align}
    \frac{\mathrm{d}\delta p_r}{\mathrm{d}t}&=-\pi_0\delta r-\rho_0\delta p_\phi,\qquad \frac{\mathrm{d} p_\phi}{\mathrm{d}t}=0,\notag\\
    \frac{\mathrm{d} \delta r}{\mathrm{d}t}&=\sigma_0\delta p_r,\qquad \frac{\mathrm{d}\delta \omega}{\mathrm{d} t}=\rho_0\delta r+ \tau_0 \delta p_\phi,
\end{align}
where $\rho_0,\tau_0,\pi_0$ and $\sigma_0$ can be computed from the Hamiltonian \eqref{Hamitonian},
\begin{align}
\rho_0=&\frac{\partial^2 {\bar{\mathcal{H}}}}{\partial r\partial p_\Phi}\big[r_0,0,p_\phi^0\big],\quad \tau_0=\frac{\partial^2 {\bar{\mathcal{H}}}}{\partial p_\phi^2}\big[r_0,0,p_\phi^0\big],\notag\\
    \pi_0=&\frac{\partial^2 {\bar{\mathcal{H}}}}{\partial r^2}\big[r_0,0,p_\phi^0\big],\quad \sigma_0=\frac{\partial^2 {\bar{\mathcal{H}}}}{\partial p_r^2}\big[r_0,0,p_\phi^0\big].
\end{align}
Seeking perturbations proportional to $e^{i \sigma t}$, one finds stable circular orbits if and only if \cite{PhysRevD.47.3281,LucBlanchet_2003,mtv7-lkv8}
 \begin{equation}
    \hat{C}_0\equiv \pi_0\sigma_0>0.
\end{equation}
Substituting the Hamiltonian $\bar{\mathcal{H}}$ into these equations, we obtain
\begin{widetext}
\begin{align}
     \frac{(G m)^2}{x^3}\hat{C}_0=1+\frac{x}{\mathcal{Z}^{4/3}}\bigg\{-6+\mathcal{Z}\zeta_0+\frac{1}{2}\Big(15-9\mathcal{Z}+12\eta\Big)\mathcal{Q}+\frac{1}{2}(1+4\eta)(-3+\mathcal{Z})\mathcal{Q}^2\bigg\}+\mathcal{O}(c^{-4}).
\end{align}
For the ISCO, there is a critical condition $\hat{C}_0=0$ \cite{LucBlanchet_2003}, and the gauge invariant PN parameter $x_{ISCO}$ is
\begin{align}\label{OmegaISCO}
    x_{ISCO}=\mathcal{Z}^{4/3}\frac{1}{6}\bigg\{1+\frac{1}{6}\zeta_0+\Big(\frac{1}{2}+\eta\Big)\mathcal{Q}\bigg\}+\mathcal{O}(\zeta_0^{2})+\mathcal{O}(\mathcal{Q}^{2})+\mathcal{O}(c^{-4}).
\end{align}
This expression is expanded for small charge parameters, $\zeta_0 \ll 1$ and $\mathcal{Q} \ll 1$. If $m_2\ll m_1$ and $q_2=0$, we have $\mathcal{Z}= 1$ and $\mathcal{Q}=0$, which agrees with the angular frequency parameter of a neutral test particle on the RN metric as 
\begin{align}
    x^{(RN)}_{ISCO}=\frac{1}{6}\bigg\{1+\frac{1}{6}\frac{Q^2}{M^2}\bigg\}+\mathcal{O}\bigg[\Big(\frac{Q}{M}\Big)^4\bigg].
\end{align}
The above result is derived in Appendix \ref{App.C}. Using Eq. \eqref{ro} and Eq. \eqref{Omega} in the harmonic coordinates, we obtain\footnote{ Eq.\eqref{OmegaISCO}  gives the leading order result for $x_{ISCO}$ in the PN expansion, whereas Eq. \eqref{ro} includes the next-to-leading order contribution to $r$. Consequently, Eq. \eqref{rISCO}  captures only part of the next-to-leading order contribution to $r_{ISCO}$. A complete next-to-leading order calculation therefore requires further work. }
\begin{align}\label{rISCO}
    r_{ISCO}=\frac{G m}{c^2}\mathcal{Z}^{-1}\bigg\{5+\mathcal{Z} \left(\frac{\eta}{3}-\frac{4 \zeta_0}{3}\right)+\mathcal{Q}^2 \left[\left(-\frac{8 \eta}{3}-\frac{2}{3}\right) \mathcal{Z}+6 \eta+\frac{3}{2}\right]+\mathcal{Q} \left[\left(\frac{\eta}{3}+6\right) \mathcal{Z}-6 \eta-\frac{13}{2}\right]\bigg\}.
\end{align}
\end{widetext}
This PN result for $r_{ISCO}$ is associated with the Coulomb effect $q_1q_2$ and the gravity effect indicated by $q_1^2$ and $q_2^2$ without a restriction of $q\ll m$. It will return to the neutral ISCO $r_{ISCO}=(5+\eta/3)G m/c^2$ in the harmonic coordinates, when $q_1=q_2=0$. 


\begin{figure}[h!]
    \centering
    \includegraphics[width=1\linewidth]{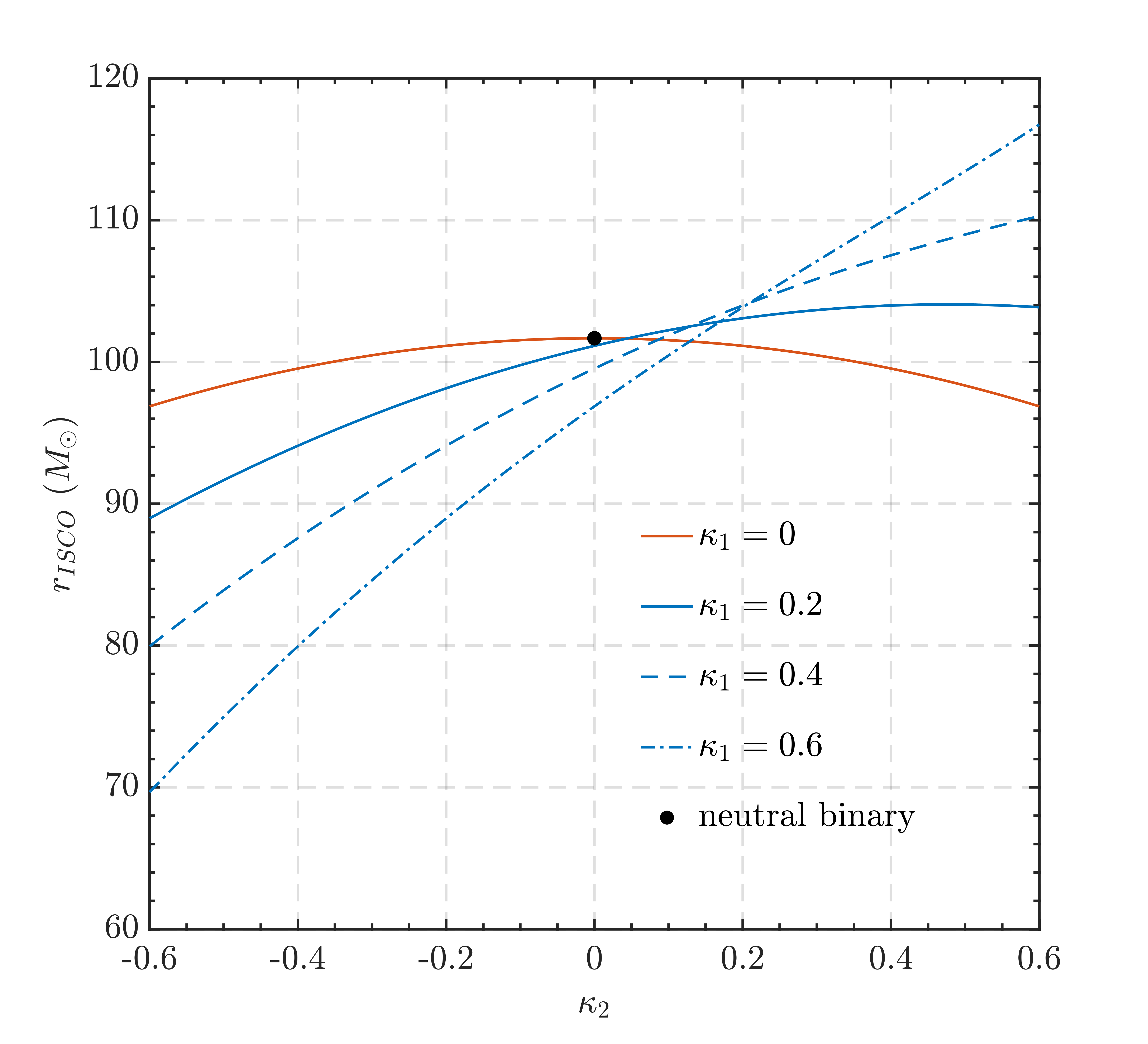}
    \caption{The PN ISCO of a charged binary in the harmonic coordinate. The black dot is the  ISCO for two neutral equal-mass  BHs of $10 M_\odot$.   {The colored curves show the ISCO radius for different charge-to-mass ratios,} $\kappa_1$ and $\kappa_2$, ranging from -0.6 to 0.6.}
    \label{RISCO}
\end{figure}
 We plot $r_{ISCO}$ with different $\kappa_1$ and $\kappa_2$ in Fig. \ref{RISCO}. The black point corresponds to the ISCO for the neutral equal-mass BHs of $10 M_\odot$. The colored curves show $r_{ISCO}$ for the charge-to-mass ratios $\kappa_1$ and $\kappa_2$ indicated in the legend. The red line represents the binary consisting of a charged BH and a neutral BH. Because the charge contributes to the gravitational field, the ISCO radius decreases as the absolute value of $\kappa_2$ increases.  The other colored lines are affected by the Coulomb interaction: the coupling factor $\mathcal{Z}$ depends on the charges. For two charges with opposite signs, $\mathcal{Q}<0$ and $\mathcal{Z}>1$, the ISCO radius becomes smaller. With two charges with the same sign, $\mathcal{Q}>0$ and $\mathcal{Z}<1$, the ISCO radius is larger. 

\subsection{Numerical results}

We numerically compute the evolution of the orbital angular frequency in Eq. \eqref{Domega} in this section. First, we need to determine the range of validity of the 1PN equation of motion. For a quasi-circular orbit, we use the dimensionless parameter $x$ as a proxy for the PN expansion parameter and impose the cutoff $x<x_{ISCO}$ which    {provides a practical cutoff for the evolution of binary orbits.}


The charge-to-mass ratios $\kappa_1$ and $\kappa_2$ are important to the equation of motion of a binary. There are four binary cases: 

\begin{enumerate}
    \item[(a)] two neutral BHs $\kappa_1=\kappa_2=0$, 
    \item[(b)] a charged BH $\kappa_1\neq0$ and a neutral BH $\kappa_2=0$, 
    \item[(c)] two charged BHs with the same sign $\kappa_1\kappa_2>0$, 
    \item[(d)] two charged BHs with opposite signs $\kappa_1\kappa_2<0$. 
\end{enumerate}

\begin{figure}[h!]
    \centering
    \includegraphics[width=1\linewidth]{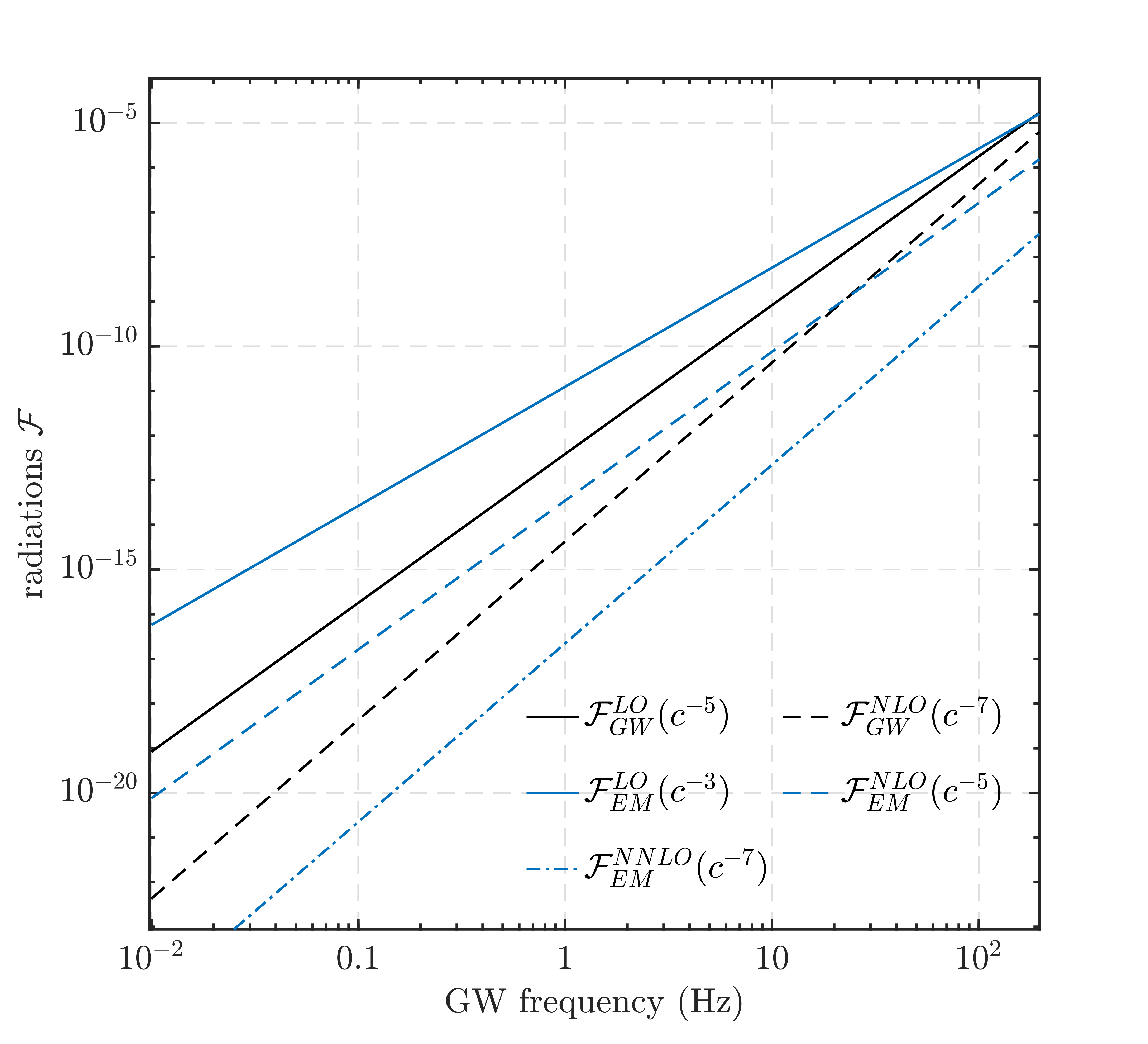}
    \caption{Contributions of different PN orders to the EM and GW radiation. We plot the leading order, NLO, and NNLO EM and GW radiation with  $m_1=10 M_\odot$, $m_2=5 M_\odot$ and $\kappa_1=-\kappa_2=0.6$. The blue lines represent EM radiation, and the black lines represent GW radiation.}
    \label{fPN}
\end{figure}

Using $f_{GW}=\omega/\pi$, we obtain the GW-frequency evolution by numerically integrating $\dot \omega$. We have the 1.5PN ($c^{-3}$) radiation from the electric dipole, the 2.5PN ($c^{-5}$) radiation from the electric dipole, quadrupole, and mass quadrupole, and the 3.5PN ($c^{-7}$) radiation from the electric and mass quadrupole and octupole. Fig. \ref{fPN} shows the contributions to the energy radiation at different PN orders. The binary system has  $m_1=10 M_\odot$, $m_2=5 M_\odot$ and $\kappa_1=-\kappa_2=0.6$ with an initial GW frequency $f_{GW0}=50$Hz.

\begin{figure}[h!]
    \centering
    \includegraphics[width=1\linewidth]{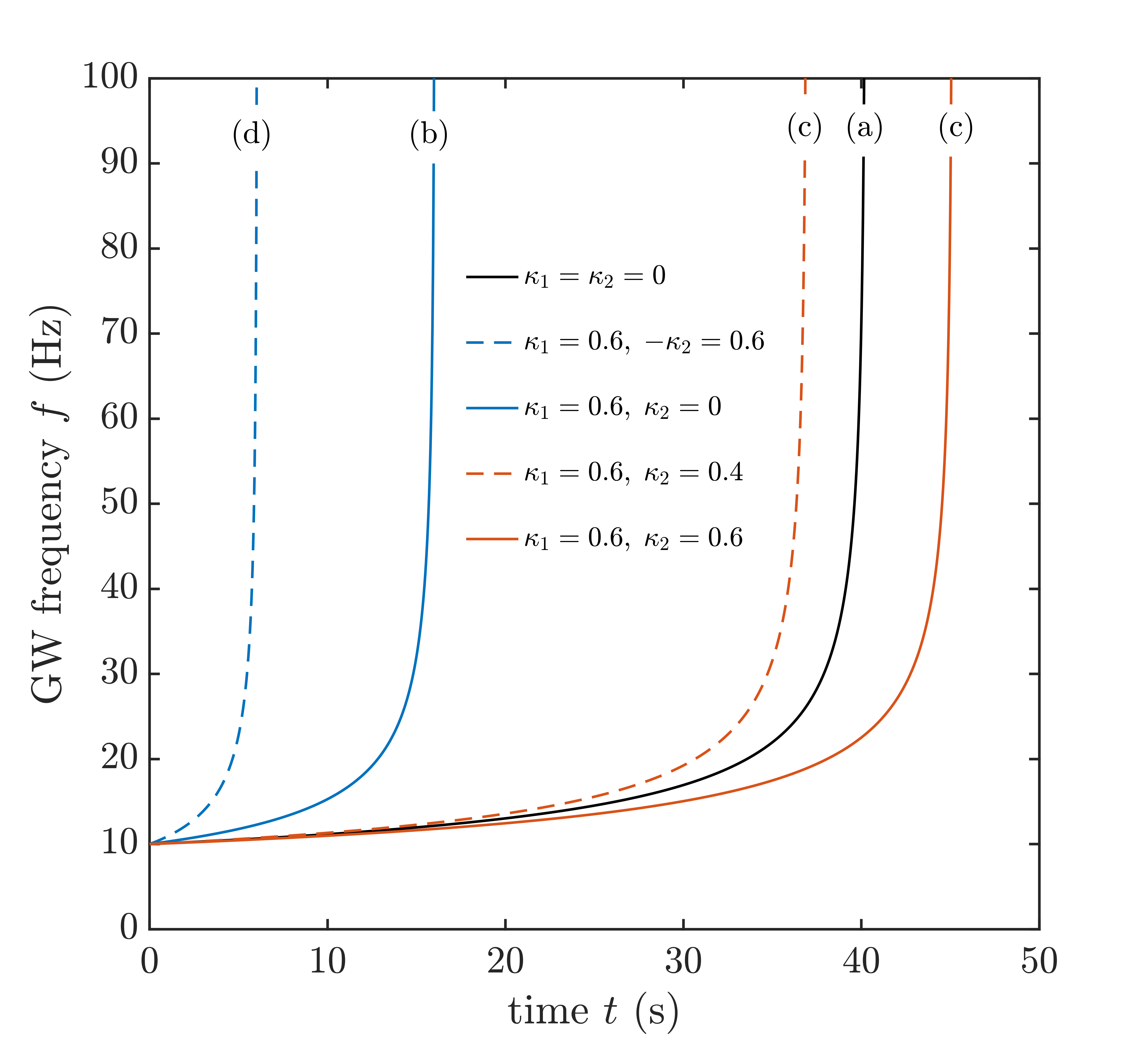}
    \caption{The evolution of the GW frequency with different charge-to-mass ratios of four cases of binaries. The black dotted line is a binary without charge and colored lines are binaries with different $\kappa_2$. The binary system has  $m_1=m_2=10 M_\odot$,   with an initial GW frequency $f_{GW0}=10$Hz.}
    \label{GWft}
\end{figure}

Fig. \ref{GWft} shows that the GW frequency of a case (d) binary increases fastest with the largest EM radiation. The case (b) binary inspirals faster than the case (a) binary because of the additional EM radiation of a single charged black hole. Case (c) is qualitatively different, charges with the same sign produce a repulsive force and lead to $\mathcal{Z}<1$. In other words, the acceleration of this type of binary is smaller than in the case (a) binary with the same orbital angular frequency. The radiation is based on the third time derivative of multipole moments, which are associated with acceleration. Although case (c) binary has additional EM radiation, the reduced acceleration can lead to a smaller total flux (EM + GW), as shown in the red solid curve inspiral more slowly than case (a) and $\dot{f}_{(a)}>\dot{f}_{(c)}$. Taking into account a symmetric situation $m_1=m_2$ and $\kappa_1=\kappa_2$, the EM dipole radiation at the leading order is strictly zero.
In other cases, where $ m_1 = m_2 $ but  $\kappa_1 \neq \kappa_2$ , the EM radiation emitted intensifies as the difference $ |\kappa_1 - \kappa_2|$ increases. Consequently, by varying the ratios  $\kappa_1 $ and  $\kappa_2 $, one can encounter regimes in which  $\dot{f}(a) < \dot{f}(c)$ .
For $\kappa_2=0.4$, the red dotted curve shows the binary inspiral in case (c) faster than in case (a) and $\dot{f}_{(a)}<\dot{f}_{(c)}$.

\begin{figure*}[ht!]
    \centering
    \includegraphics[width=0.5\linewidth]{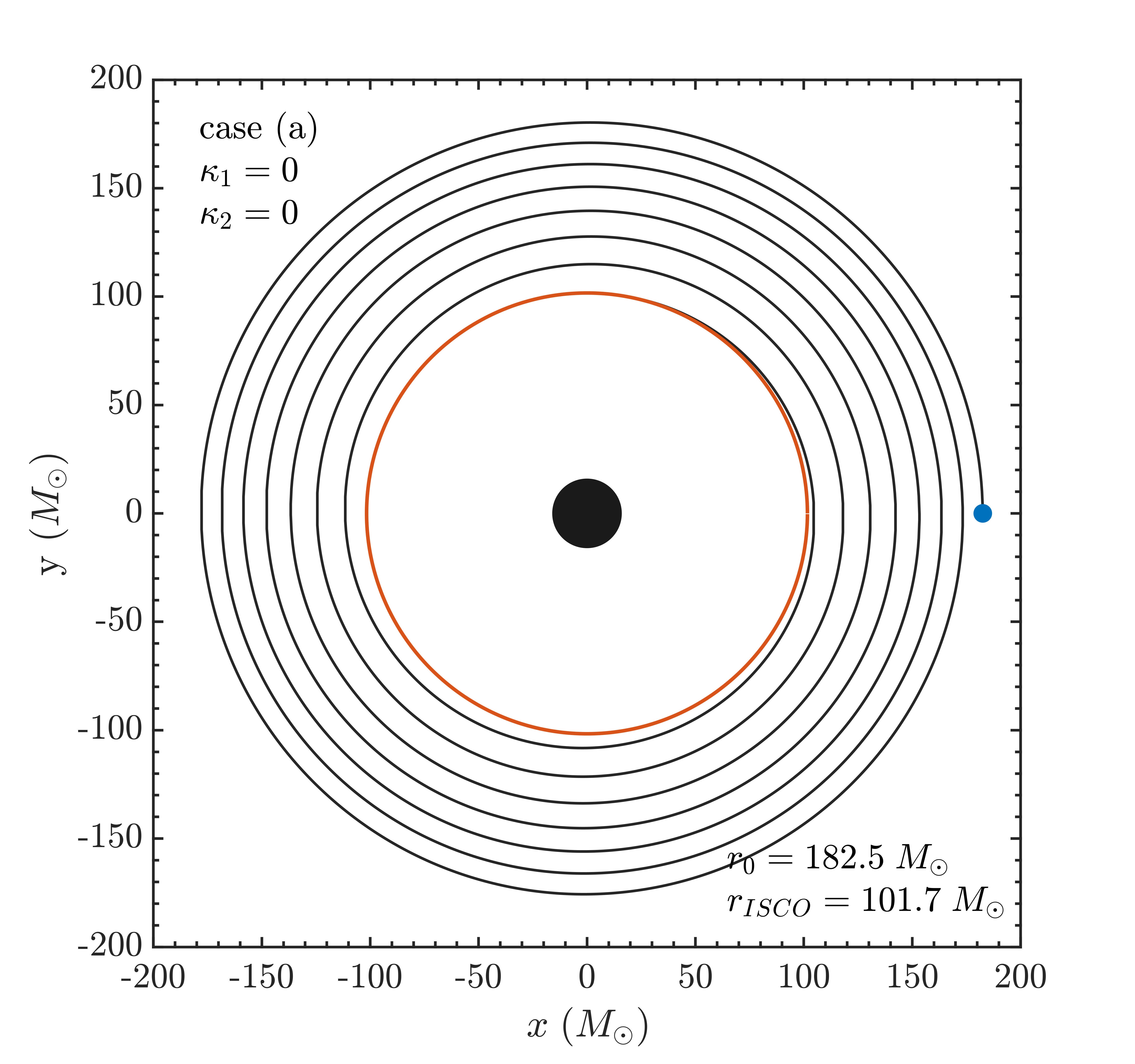}\includegraphics[width=0.5\linewidth]{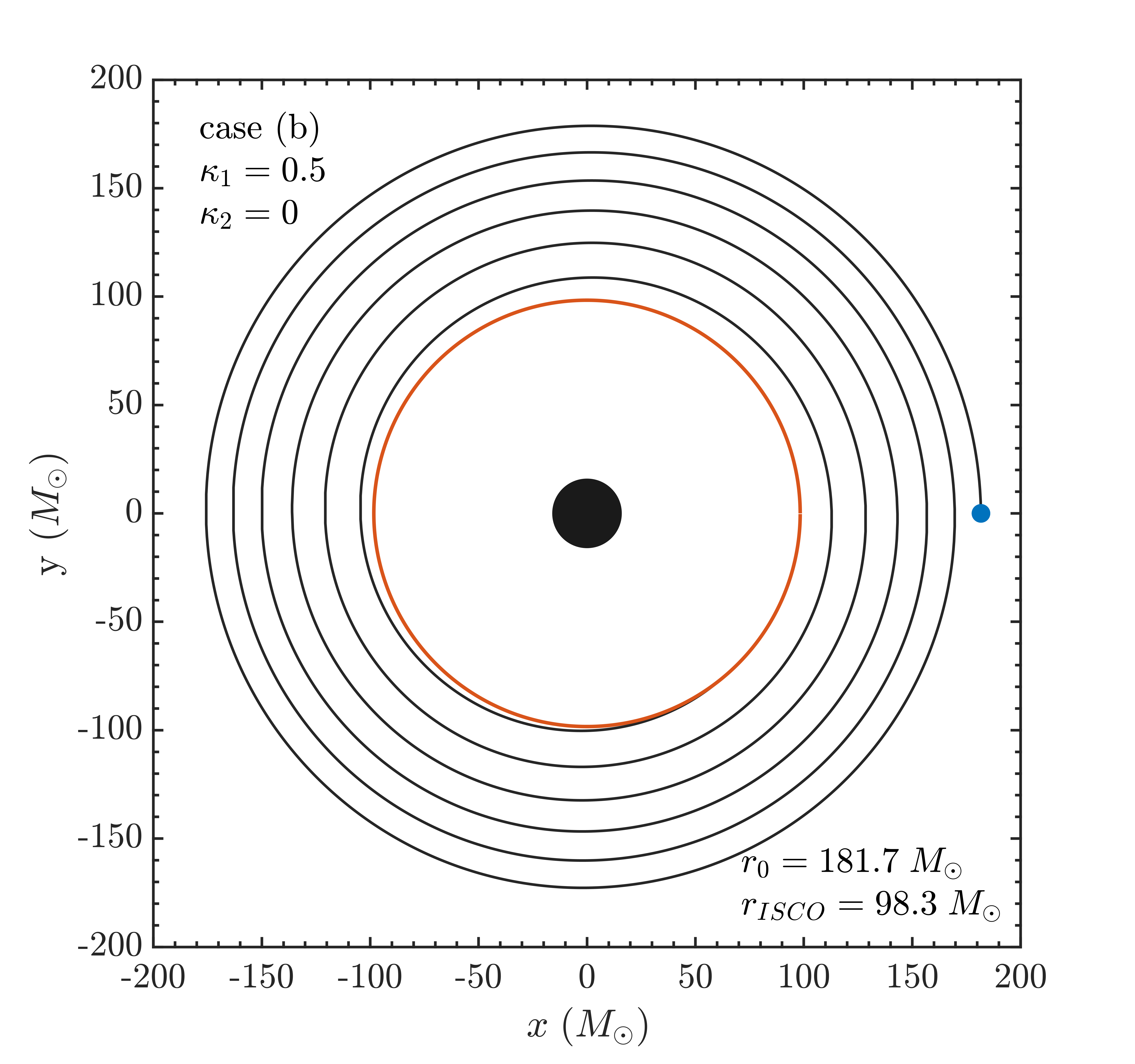}\\
    \includegraphics[width=0.5\linewidth]{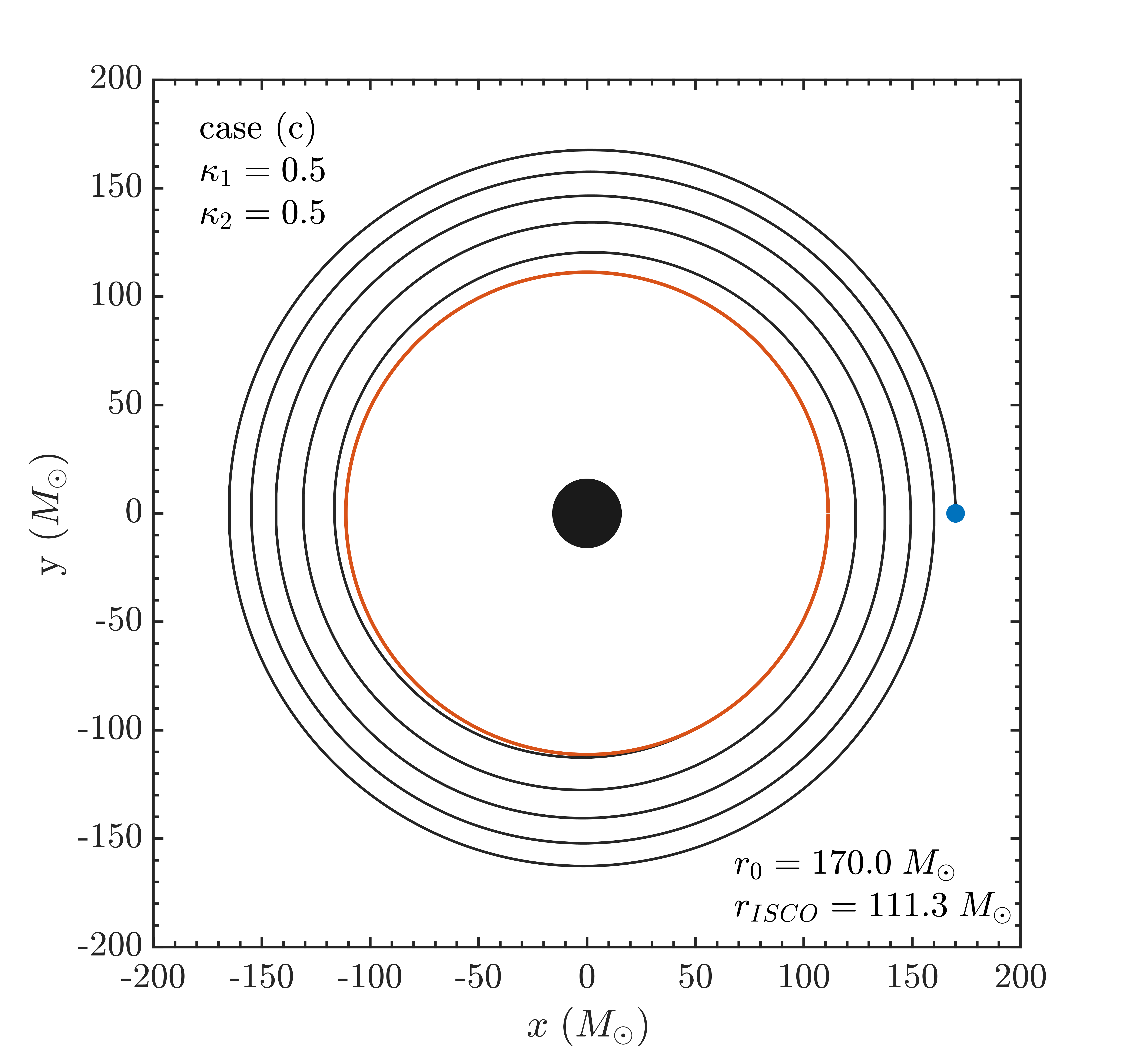}\includegraphics[width=0.5\linewidth]{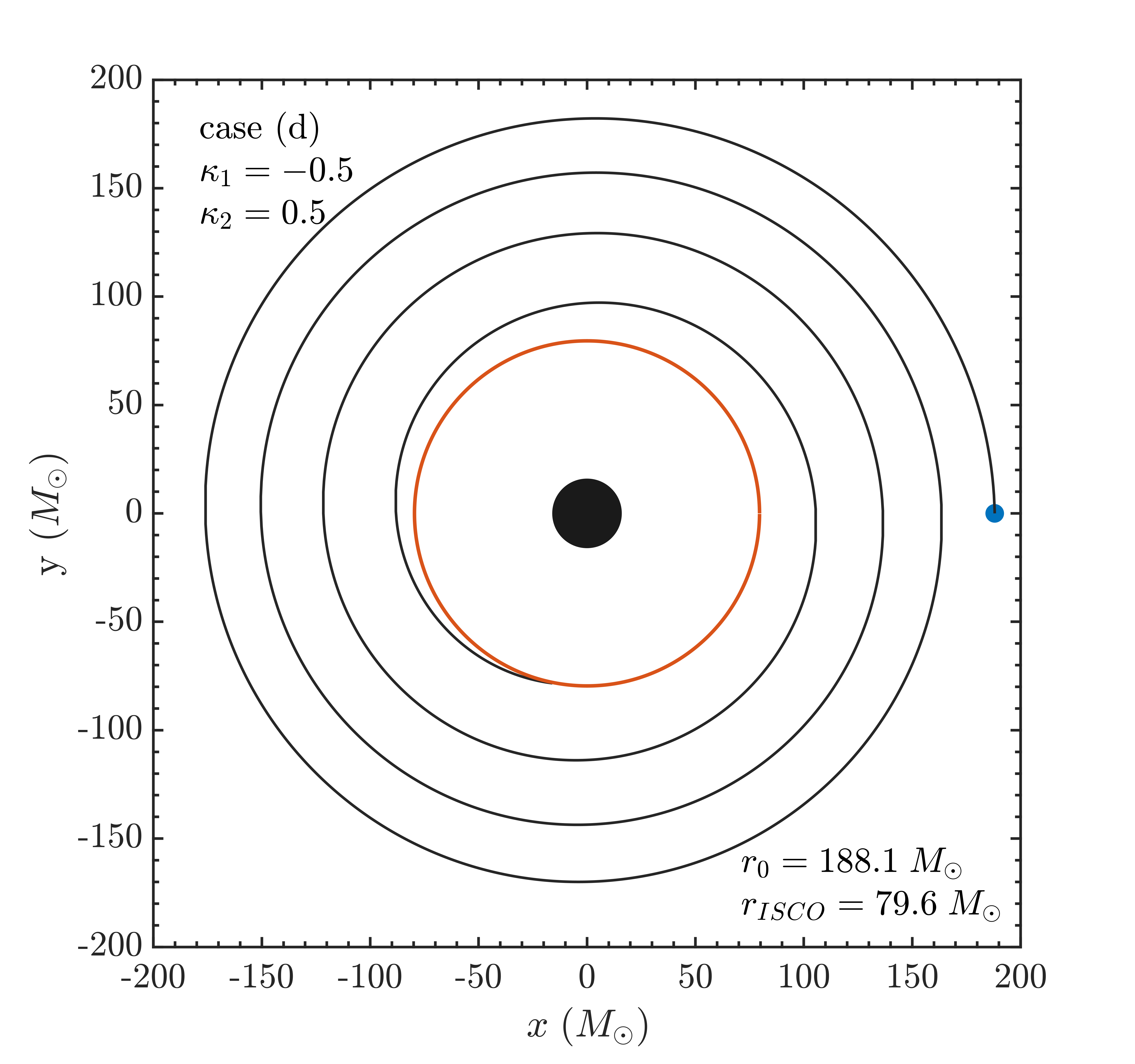}
    \caption{Orbital trajectories for the four representative cases. Red dashed circles mark the corresponding ISCO radii. For a binary $m_1=m_2=10 ~M_\odot$ with the initial GW frequency $ f=100$ Hz, different charge-to-mass ratios $\kappa_1$ and $\kappa_2$ led to different orbital trajectories.}
    \label{Orbits}
\end{figure*}

We plot the typical trajectories of four binary cases in Fig. \ref{Orbits}. The red circles denote ISCOs; if the separation of a binary is smaller than that of ISCO, the PN method is not valid. We cut off the trajectories with the condition $x<x_{ISCO}$. It can be seen that the orbit of the case (d) binary shrinks much faseter than the case (a) binary.

\section{Conclusion}\label{Sec5}
Electromagnetic effects can play an important role in the dynamics of charged BBHs. In this work, we rederive the conservative 1PN acceleration in the center-of-mass frame in Eq. \eqref{aCoM}. At high orbital angular frequency, charged binaries can produce significant EM and GW fluxes, which significantly affect the evolution of the orbital angular frequency. We specifically derive the EM and GW radiation arising from variations in electric and mass multipole moments. Using the leading-order MPM matching and the 1PN source-multipole expansion, we obtain the source multipole moments given in Eq. \eqref{Multipoles}. Because the conservative 1PN acceleration contains EM terms, GW radiation is also affected by EM effects. Using the flux-balance equations, we obtain the evolution of the gauge-invariant orbital angular frequency within quasi-circular orbits due to the back-reaction of EM and GW radiation. The conservative ingredients entering our flux-balance analysis have been derived through 1PN order. 
We have fully derived the GW and EX fluxes up to NLO and have partially obtained the NNLO EM flux. Since the leading order EM flux enters at \(-1\) PN relative to the GW flux, a complete 1PN treatment of the dissipative effects would require the full NNLO EM flux. We leave this completion for future work. The gauge invariant analytical PN parameter $x$ at ISCO in the PN framework is given in Eq. \eqref{OmegaISCO}, which is consistent with the ISCO of a neutral test particle in the RN metric.

The dynamics of BBHs are significantly affected by the charge-to-mass ratios and the signs of charges. We divide  BBHs into four representative cases: (a) two neutral BHs $\kappa_1=\kappa_2=0$, (b) a charged BH and a neutral BH $\kappa_1\neq0$, $\kappa_2=0$, (c) two charged BHs with the same sign $\kappa_1\kappa_2>0$, (d) two charged BHs with opposite signs $\kappa_1\kappa_2<0$. We compute the orbital trajectories for each type of BBHs and discuss the effects of the EM radiation: case (d) binaries will speed up the process from inspiral to merger; case (c) binaries are qualitatively distinct, while the speed of inspiral is determined by the charge-to-mass ratios of BBHs.

It is interesting to study the EM and GW radiation  back-reaction acceleration in different gauges \cite{Gravity_PoissonWill}. For GW astronomy, we already have the evolution of the GW-frequency evolution at the stated radiation-sector accuracy, which is the most important parameter for GW detection. A natural next step is to develop waveform templates that include EM effects and use them in GW data analysis, which would allow us to obtain more information about BBH sources.

\section{ACKNOWLEDGMENTS}
T. L. is supported by the China Postdoctoral Science Foundation Grant No. 2024M760692. This work is supported in part by the National Natural Science Foundation of China under Grant No. 12475067 and No. 12235019.

\newpage
\appendix

\section{The post-Newtonian potentials}\label{App.A}
In this appendix, we derive the EM and gravitational PN potentials in Eq. \eqref{PNpotentials}. For the EM sector, varying $S_EM$ with respect to $A_mu$ yields the leading-order (LO) equation of motion
\begin{equation}\label{AJ}
    \Box A^{\text{LO}}_\mu=-\sum_{A=1,2} \frac{4\pi G  }{c^3}\eta_{\mu\nu} \delta_A q_A v_A^\nu+\mathcal{O}(c^{-4}),
\end{equation}
where we retain the Newtonian-order contribution to the EM source $\iota$. We use Hadamard partie-finie regularization to regularize the integrand $\Box^{-1}_{\text{R}}$ \cite{Blanchet_2000,661cfda7-9a69-35d7-8947-df9a1ce31bd5}
\begin{equation}
(\Box^{-1}_R F)(\bm{x}',t'):=-\frac{1}{4\pi}\underset{s}{\text{Pf}}\int \frac{\mathrm{d}^3\bm{x}}{|\bm{x}-\bm{x}'|}F\Big(\bm{x},t'-\frac{1}{c}|\bm{x}'-\bm{x}|\Big),
\end{equation}
where we denote the spherical ball $r=|\bm{x}'-\bm{x}| < s$ of radius $s$ and centered on the singularity by $\mathcal{B}(s)$. Retaining the finite terms in Eq. \eqref{AJ} which integrate over a domain $\mathbb{R}^3 \setminus \mathcal{B}(s)$, and expanding in multipoles
\begin{equation}
    \frac{1}{|\bm{x}-\bm{x}'|}=\frac{1}{r}+\frac{\bm{n}\cdot\bm{x}'}{r^2}+\mathcal{O}(\frac{1}{r^3}),
\end{equation}
we obtain the field $P$ at the position of particle $A$ 
\begin{align}
    P|_A=&-\frac{1}{4\pi}\underset{s}{\text{Pf}}\int \mathrm{d}^3x\frac{1}{r}F(x)+\frac{\bm{n}\cdot\bm{x}'}{r^2}F(x)\notag\\
    &+\left[\ln\left(\frac{r'}{s}-1\right)\Big(r^2F(x)\Big)_A\right].
\end{align}
After these operations, we can eliminate the divergent terms of integration and obtain $A_\mu^\text{LO}$ as the usual formula for constant dipoles \cite{Jackson:1998nia}
\begin{align}
\varphi|_A=&\frac{q_B}{r}+\mathcal{O}(c^{-2}),\\
\chi_i|_A=&-\frac{1}{2}\bigg(\frac{q_B v_B^i}{r}+\frac{I_B^{ik} n_k}{r^2}\bigg)+\mathcal{O}(c^{-2}),
\end{align}
where $q_A^k=q_A x^k$ is the electric dipole of particle A,
\begin{equation}
I^{ik}_A=\epsilon_{ijk}\mu^k_A=\frac{1}{2}( x_A^i j^k_A-x_A^k j^i_A),
\end{equation}
and $\mu^k_A$ is the magnetic dipole of particle A.
To match the RN metric, $A_0$ is of the order of $c^{-1}$ while $A_i$ is of the order of $c^{-2}$. This rescaling is useful for the PN counting used in the following.

The NLO of the equation of motion of $A_0$ is
\begin{align}
        \Box A^{\text{NLO}}_0&=-\sum_{A=1,2} \frac{4\pi G  }{c^3}g_{0\nu} \delta_A q_A v_A^\nu+\Phi_0\notag\\
        &=-\sum_{A=1,2} \frac{4\pi G  }{c^3}g_{0\nu} \delta_A q_A v_A^\nu-\frac{2}{c^2}\partial_i V \partial_i A^{\text{LO}}_0\notag\\
        &=\sum_{A=1,2} \frac{4\pi G  }{c^2} \delta_A q_A-\frac{2}{c^2}\big(V \nabla^2 A^{\text{LO}}_0 + \partial_i V \partial_i A^{\text{LO}}_0\big)\notag\\
        &=\sum_{A=1,2} \frac{4\pi G  }{c^2} \delta_A q_A\notag\\
        &+\frac{1}{c^2}\Big[A^{\text{LO}}_0 \nabla^2 V -  V \nabla^2 A^{\text{LO}}_0-\nabla^2(VA^{\text{LO}}_0)\Big],
\end{align}
where $\Phi_0$ is the non-linear part of the effective EM source in Eq. \eqref{Phi}. Substituting the leading order (Newtonian order) expressions for $A_0^{\rm LO}$ and $V$, we set the 1PN EM potential
\begin{align}
  \varphi|_1=&\frac{q_2}{r}+\frac{1}{c^2}
\Box^{-1}(A^{\text{LO}}_0 \nabla^2 V -  V \nabla^2 A^{\text{LO}}_0)+\mathcal{O}(c^{-4})\notag\\
=&\frac{q_2}{r}+\frac{1}{c^2}\underset{s}{\mathrm{Pf}} \int \mathrm{d}^3 \mathbf{x}  \frac{4\pi G \delta_1}{|\bm{x}-\bm{x}_2|}(m_2q_1-m_1q_2)\notag\\
&+\mathcal{O}(c^{-4})\notag\\
=&\frac{q_2}{r}+\frac{G}{c^2r^2}(m_2q_1-m_1q_2)+\mathcal{O}(c^{-4}).
\end{align}
  {Performing the partie-finie integral to the $\nabla^2(VA^{\text{LO}}_0)$ term and discarding terms that vanish in the limit \cite{PhysRevD.58.124002,placidi2025chargedblackholebinaryevolution}.
Returning to the 1PN electric potential $\varphi$, we obtain the 1PN EM vector 
\begin{align}
    A_0&=\frac{1}{c}\varphi-\frac{1}{c^3}V\varphi+\mathcal{O}(c^{-5}),\\
    A_i&=\frac{1}{c^2}\chi_i+\mathcal{O}(c^{-4}).
\end{align}
}
  {Returning to the field equation \eqref{FieldEQ}, non-compact parts $\Lambda_{\mu\nu}\thicksim \mathcal{O}(c^{-4})$ match to the NNLO of $T_{\mu\nu}$ while $\Phi_{\mu}\thicksim \mathcal{O}(c^{-4})$ match to the NLO of $A_{\mu}$. So that we can calculate $T_{\mu\nu}$ without the contributions of $\Lambda_{\mu\nu}$, while the EM source $\iota_{1}^\mu$ of charge A at 1PN order need to consider the non-linear contributions of the effective EM source $\Phi_\mu|_1$ in Eq. \eqref{iota}. We obtain
\begin{align}
    \iota_{1}^0&=\sqrt{-g}j^0_1+g^{00}\frac{2c^2}{4\pi}\partial_i V|_1\partial_i A_0|_2\notag\\
    &=c\delta_1q_1- \frac{G}{4\pi c}\partial^2_i\partial^2_i\frac{ m_2 q_1}{|\bm{x}-\bm{x}_1||\bm{x}-\bm{x}_2|}\notag\\
    &+\mathcal{O}(c^{-3}),\\
    \iota_{1}^i&=\sqrt{-g}j^i_1+\delta^{ij}\frac{2c^2}{4\pi}\partial_k V|_1\partial_k A_j|_2\notag\\
    &=\delta_1q_1v_1^i- \frac{G}{8\pi c^2}\partial^2_k\partial^1_k\frac{ m_2 q_1 \Big[(\bm{v}_1\cdot\bm{n})n^{i}+v_{1}^{i}\Big]}{|\bm{x}-\bm{x}_1||\bm{x}-\bm{x}_2|}\notag\\
    &+\mathcal{O}(c^{-4}).
\end{align}
Performing the partie-finie integral to $\iota^\mu_A$, we obtain
\begin{align}
    \underset{s}{\text{Pf}}\int\mathrm{d}^3\mathbf{x}\iota_{1}^0&=c q_1- \frac{G  m_2 q_1}{cr}+\mathcal{O}(c^{-3}),\\
    \underset{s}{\text{Pf}}\int\mathrm{d}^3\mathbf{x}\iota_{1}^i&=q_1 v^i_1- \frac{G  m_2 q_1}{c^2r}\Big[(\bm{v}_1\cdot\bm{n})n^{i}+v_{1}^{i}\Big]\notag\\
    &+\mathcal{O}(c^{-4}),
\end{align}
which is useful for calculating the electric multipole moments in the following section.
}

For gravitational PN potentials, we associate them with Newtonian potentials as the standard PN process \cite{Blanchet2024}
\begin{align}
    V&=U+\frac{1}{c^2}\frac{1}{2}\partial_t^2(2\bigtriangleup^{-1}U)+\mathcal{O}(c^{-4}),\\
    V_i&=U_i+\mathcal{O}(c^{-2}),
\end{align}
and $U$ can be expressed by the stress-energy tensor
\begin{align}\label{UT}
U(\bm{x},t)&=G\int\frac{\mathrm{d}^3 \bm{x}'}{|\bm{x}-\bm{x}'|}\left[T^{00}(\bm{x}',t)+T^{ii}(\bm{x}',t)\right],\\
    U^i(\bm{x},t)&=G\int\frac{\mathrm{d}^3 \bm{x}'}{|\bm{x}-\bm{x}'|}\left[T^{0i}(\bm{x}',t)\right].
\end{align}
The stress-energy tensor of the point particle in Eq.~\eqref{Tmunu} can be expanded as

\begin{align}
    T^{00}_{\rm pp}=&\sum_{A=1,2} m_Ac^2\delta_A\bigg[1+\frac{1}{c^2}\Big(\frac{1}{2}v_A^2-U_A\Big)\bigg]+\mathcal{O}(c^{-2}),\\
T^{0i}_{\rm pp}=&\sum_{A=1,2} m_Ac\delta_Av_A^i+\mathcal{O}(c^{-1}),\\
T^{ij}_{\rm pp}=&\sum_{A=1,2} m_A\delta_Av_A^iv_A^j+\mathcal{O}(c^{-2}).
\end{align}
We perform a PN iteration between the PN potential and the stress-energy tensor and obtain
\begin{align}
    V|_1&=\frac{Gm_2}{r}\bigg\{1+\frac{1}{c^2}\bigg[\frac{3}{2}v_1^2-\frac{Gm_1}{r}+\frac{1}{2}\bm{v}_1\cdot\bm{v}_2\notag\\
    &+\frac{1}{2}(\bm{v}_1\cdot\bm{n})(\bm{v}_{2}\cdot\bm{n})\bigg]\bigg\}+\mathcal{O}(c^{-4}),\\
    V_i|_{1}&=\frac{Gm_2}{r}v_{2i}+\mathcal{O}(c^{-2}).
\end{align}
Using a total time derivative, we can eliminate the acceleration terms at the 1PN potential $V|_A$
\begin{equation}
    \partial_t(\bm{v}\cdot \bm{n})=\bm{a}\cdot\bm{n}+\frac{1}{|\bm{x}|}\left[v^2-(\bm{v}\cdot\bm{n})^2\right].
\end{equation}

We have now obtained all post-Newtonian potentials. By substituting them into the total action $S$ in Eq. \eqref{action}, we derive the Fokker action. This result contains only positions and velocities, as shown in Eq. \eqref{Lagrangian}.

\begin{widetext}
\section{Multipole expansions of radiation}\label{App.B}

In this appendix, we derive the electric and mass multipole moments we have used in Eq. \eqref{Multipoles}, as well as the electric and gravitational radiation to 3.5PN order. We use the STF expansion form of tensors and vectors as \cite{RevModPhys.52.299,PhysRevD.85.125033}
\begin{align}
    \text{F}^{\mu\nu}_ L&=\underset{B=0}{\text{FP}}\int \mathrm{d}^3 x \Big(\frac{r}{r_0}\Big)^B\hat{x}_L\int^1_{-1} \mathrm{d} z\delta_\ell (z)\tau^{\mu\nu}(\bm{x},t+zr/c),\\
    \text{G}^{\mu}_{ L}&=\underset{B=0}{\text{FP}}\int \mathrm{d}^3 x \Big(\frac{r}{r_0}\Big)^B\hat{x}_L\int^1_{-1} \mathrm{d} z\delta_\ell (z)\iota^\mu(\bm{x},t+zr/c),
\end{align}
  {where the   source in Eq. \eqref{Fieldh} is 
\begin{gather}
\tau^{\mu\nu}=T^{\mu\nu}_{\text{pp}}+\Lambda^{\mu\nu}_{EM},\label{tau}\\
\Lambda^{00}_{EM}=\frac{\epsilon_0}{2}(\partial_k \varphi)^2+\mathcal{O}(c^{-2}),\qquad \Lambda^{ij}_{EM}=\frac{\epsilon_0}{2}\Big[\delta^{ij}(\partial_k \varphi)^2-2\partial^i \varphi\partial^j \varphi\Big]+\mathcal{O}(c^{-2}),\qquad \Lambda^{0i}_{EM}=\mathcal{O}(c^{-2}).
\end{gather}}
$(r/r_0)^B$ is the divergent term of the integration $\underset{B=0}{\text{PF}}$, which removes the divergence coming from $1/|\bm{x}-\bm{x}_B|$ as shown in the last section. The z-integration involves the weighting function 
\begin{equation}
    \delta_\ell(z)=\frac{(2\ell+1)!!}{2^{\ell+1}\ell !}(1-z^2)^\ell,
\end{equation}
the gravitational field $h_{\mu\nu}$ and the EM field $A_\mu$ can be written as
\begin{align}
    h^{\mu\nu}=-\frac{4G}{c^4}\sum_{\ell\geq0}\frac{(-)^\ell}{\ell !}\partial_L\bigg[\frac{1}{r}F^{\mu\nu}_L(t-r/c)\bigg],\\
    A_{\mu}=\frac{4G}{c^4}\sum_{\ell\geq0}\frac{(-)^\ell}{\ell !}\partial_L\bigg[\frac{1}{r}G_{\mu L}(t-r/c)\bigg].
\end{align}
The energy and angular momentum radiation are
\begin{align}
\mathcal{F}_{GW}=&\sum^{+\infty}_{\ell=2}\frac{G}{c^{2\ell+1}}\frac{(\ell+1)(\ell+2)}{(\ell-1)\ell\ell!(2\ell+1)!!}\bigg[\text{U}^{(1)}_{L}\text{U}^{(1)}_{L}+\frac{4\ell^2}{c^2(\ell+1)^2}\text{V}^{(1)}_L\text{V}^{(1)}_L\bigg],\\
\mathcal{G}^i_{GW}=&\sum^{+\infty}_{\ell=2}\frac{G}{c^{2\ell+1}}\frac{(\ell+1)(\ell+2)}{(\ell-1)\ell!(2\ell+1)!!}\bigg[\text{U}_{jL-1}\text{U}^{(1)}_{kL-1}+\frac{4\ell^2}{c^2(\ell+1)^2}\text{V}_{jL-1}\text{V}^{(1)}_{kL-1}\bigg]\epsilon_{ijk},\\
\mathcal{F}_{EM}=&\sum^{+\infty}_{\ell=1}\frac{G}{c^{2\ell+1}}\frac{\ell+1}{\ell\ell!(2\ell+1)!!}\bigg[\text{Q}^{(1)}_{L}\text{Q}^{(1)}_{L}+\frac{\ell^2}{c^2(\ell+1)^2}\text{H}^{(1)}_L\text{H}^{(1)}_L\bigg],\\
\mathcal{G}^i_{EM}=&\sum^{+\infty}_{\ell=1}\frac{G}{c^{2\ell+1}}\frac{\ell+1}{\ell!(2\ell+1)!!}\bigg[\text{Q}_{jL-1}\text{Q}^{(1)}_{kL-1}+\frac{\ell^2}{c^2(\ell+1)^2}\text{H}_{jL-1}\text{H}^{(1)}_{kL-1}\bigg]\epsilon_{ijk},
\end{align}
We match the radiation multipoles with the source multipoles at the leading order of the post-Minkowski expansion\footnote{The tail terms are ignored.},
\begin{equation}
     \mathrm{U}_L=\mathrm{I}^{(\ell)}_L+\mathcal{O}(G),\quad \mathrm{V}_L=\mathrm{J}^{(\ell)}_L+\mathcal{O}(G),\quad\mathrm{Q}_L=\mathrm{E}^{(\ell)}_L+\mathcal{O}(G),\quad \mathrm{H}_L=\mathrm{B}^{(\ell)}_L+\mathcal{O}(G).
\end{equation}
Matching $\mathrm{U}_L,\mathrm{V}_L,\mathrm{Q}_L,\mathrm{H}_L$ to the mass and electric multipoles, we obtain
\begin{align}
\mathcal{F}_{GW}=&\frac{G}{c^5}\frac{1}{5}\big\langle \hat{\mathrm{I}}^{(3)}_{ij}\hat{\mathrm{I}}^{(3)}_{ij}\big\rangle+\frac{G}{c^7}\bigg[\frac{16}{45}\big\langle \hat{\mathrm{J}}^{(3)}_{ij}\hat{\mathrm{J}}^{(3)}_{ij}\big\rangle+\frac{1}{189}\big\langle \hat{\mathrm{I}}^{(4)}_{ijk}\hat{\mathrm{I}}^{(4)}_{ijk}\big\rangle\bigg]+\mathcal{O}(c^{-9}),\\
\mathcal{G}^i_{GW}=&\frac{G}{c^5}\frac{2}{5}\big\langle \hat{\mathrm{I}}^{(2)}_{jl}\hat{\mathrm{I}}^{(3)}_{kl}\big\rangle\epsilon_{ijk}+\frac{G}{c^7}\bigg[\frac{32}{45}\big\langle \hat{\mathrm{J}}^{(2)}_{jl}\hat{\mathrm{J}}^{(3)}_{kl}\big\rangle+\frac{1}{63}\big\langle \hat{\mathrm{I}}^{(3)}_{jlm}\hat{\mathrm{I}}^{(4)}_{klm}\big\rangle\bigg]\epsilon_{ijk}+\mathcal{O}(c^{-9}),\\
\mathcal{F}_{EM}=&\frac{G}{c^3}\frac{2}{3}\big\langle {\mathrm{E}}^{(2)}_{i}{\mathrm{E}}^{(2)}_{i}\big\rangle
+\frac{G}{c^5}\bigg[\frac{1}{6}\big\langle \hat{\mathrm{B}}^{(2)}_{i}\hat{\mathrm{B}}^{(2)}_{i}\big\rangle+\frac{1}{20}\big\langle \hat{\mathrm{E}}^{(3)}_{ij}\hat{\mathrm{E}}^{(3)}_{ij}\big\rangle\bigg]
+\frac{G}{c^7}\bigg[\frac{1}{45}\big\langle \hat{\mathrm{B}}^{(3)}_{ij}\hat{\mathrm{B}}^{(3)}_{ij}\big\rangle+\frac{2}{945}\big\langle \hat{\mathrm{E}}^{(4)}_{ijk}\hat{\mathrm{E}}^{(4)}_{ijk}\big\rangle\bigg]\notag\\
&+\mathcal{O}(c^{-9}),\label{FEM}\\
\mathcal{G}^i_{EM}=&\frac{G}{c^3}\frac{2}{3}\big\langle {\mathrm{E}}^{(1)}_{j}{\mathrm{E}}^{(2)}_{k}\big\rangle\epsilon_{ijk}
+\frac{G}{c^5}\bigg[\frac{1}{6}\big\langle \hat{\mathrm{B}}^{(1)}_{j}\hat{\mathrm{B}}^{(2)}_{k}\big\rangle+\frac{1}{10}\big\langle \hat{\mathrm{E}}^{(2)}_{jl}\hat{\mathrm{E}}^{(3)}_{kl}\big\rangle\bigg]\epsilon_{ijk}+\frac{G}{c^7}\bigg[\frac{2}{45}\big\langle \hat{\mathrm{B}}^{(2)}_{jl}\hat{\mathrm{B}}^{(3)}_{kl}\big\rangle+\frac{2}{315}\big\langle \hat{\mathrm{E}}^{(3)}_{jlm}\hat{\mathrm{E}}^{(4)}_{klm}\big\rangle\bigg]\epsilon_{ijk}\notag\\
&+\mathcal{O}(c^{-9}),\label{NEM}
\end{align}
where $\mathrm{I}_{ij},\mathrm{I}_{ijk}$ and $\mathrm{J}_{ij}$ are the mass quadrupole, the mass octupole, and the mass current quadrupole. Similarly, $\mathrm{E}_{i},\mathrm{E}_{ij},\mathrm{E}_{ijk},\mathrm{B}_{i}$ and $\mathrm{B}_{ij}$ are the electric dipole, the electric quadrupole, the electric octupole, the electric current dipole, and the electric current quadrupole. All hatted quantities ``$\hat{\quad}$'' are the STF form. We can use the following relationship to obtain the STF tensor
\begin{gather}
    \hat{\mathrm{A}}_{ij}={\mathrm{A}}_{\langle ij\rangle}=\mathrm{A}_{(ij)}-\frac{1}{3}\delta_{ij}\mathrm{A}_{kk},\quad \hat{\mathrm{A}}_{ijk}={\mathrm{A}}_{\langle ijk\rangle}=\mathrm{A}_{(ijk)}-\frac{1}{5}\Big(\delta_{ij}\mathrm{A}_{llk}+\delta_{ik}\mathrm{A}_{ljl}+\delta_{jk}\mathrm{A}_{ill}\Big).
\end{gather}
Here $\mathrm{A}$ can be replaced by multipoles $\mathrm{I}$, $\mathrm{E}$, $\mathrm{J}$ and $\mathrm{B}$.

We need to expand the mass quadrupole, electric dipole, and electric quadrupole to the 1PN order, which can be written as \cite{PhysRevD.109.084048,PhysRevD.85.125033}
\begin{align}
    \mathrm{I}_L=& \underset{B=0}{\mathrm{FP}} \int \mathrm{d}^3\mathbf{x} \left(\frac{r}{r_0} \right)^B \int_{-1}^{1} \mathrm{d}z \bigg\{\delta_{\ell}(z) \hat{x}_L \bar{\Sigma}- \frac{1}{c^2}\frac{4(2\ell + 1)}{(\ell + 1)(2\ell + 3)} \delta_{\ell+1}(z) \hat{x}_{iL} \bar{\Sigma}_i^{(1)}\bigg\}+\mathcal{O}(c^{-4}), \\
    \mathrm{J}_L=& \underset{B=0}{\mathrm{FP}} \int \mathrm{d}^3\mathbf{x} \left(\frac{r}{r_0} \right)^B \int_{-1}^{1} \mathrm{d}z \epsilon_{ij\langle i_\ell}\bigg\{\delta_{\ell}(z) \hat{x}_{L-1\rangle i} \bar{\Sigma}_j- \frac{1}{c^2}\frac{(2\ell + 1)}{(\ell + 2)(2\ell + 3)} \delta_{\ell+1}(z) \hat{x}_{L-1\rangle ik} \bar{\Sigma}_{jk}^{(1)}\bigg\}, \\
    \mathrm{E}_L =& \underset{B=0}{\mathrm{FP}} \int \mathrm{d}^3\mathbf{x} \left(\frac{r}{r_0} \right)^B \int_{-1}^{1} \mathrm{d}z \bigg\{\frac{1}{c}\delta_{\ell}(z) \hat{x}_L \bar{\iota}^0+ \frac{1}{c^2}\frac{2\ell + 1}{(\ell + 1)(2\ell + 3)} \delta_{\ell+1}(z) \hat{x}_{iL} \bar{\iota}^{i(1)}\bigg\}+\mathcal{O}(c^{-4}),\\
    \mathrm{B}_L =& \underset{B=0}{\mathrm{FP}} \int \mathrm{d}^3\mathbf{x} \left(\frac{r}{r_0} \right)^B \int_{-1}^{1} \mathrm{d}z \epsilon_{ij\langle i_\ell}\bigg\{\delta_{\ell}(z) \hat{x}_{L-1\rangle i} \bar{\iota}^j\bigg\}.
\end{align}
The integrals over$\delta_l(z)$ are
\begin{align}
    \int^1_{-1} \mathrm{d} z\delta_\ell (z)\bar{\iota}_{\mu}(\bm{x},t+zr/c)&=\sum^{+\infty}_{k=0}\frac{(2\ell +1)!!}{2^k k!(2\ell+2k+1)!!}\bigg(\frac{r}{c}\frac{\partial}{\partial t}\bigg)^{2k}{\iota}_\mu(\bm{x},t),\\
    \int^1_{-1} \mathrm{d} z\delta_\ell (z)\bar{\Sigma}(\bm{x},t+zr/c)&=\sum^{+\infty}_{k=0}\frac{(2\ell +1)!!}{2^k k!(2\ell+2k+1)!!}\bigg(\frac{r}{c}\frac{\partial}{\partial t}\bigg)^{2k}{\Sigma}(\bm{x},t).
\end{align}
The pure EM part is consistent with Refs. \cite{PhysRevD.43.3259,1989MNRAS.239..845B}. 
  {\begin{align}
    \hat{\mathrm{E}}^{i}=&\frac{1}{c}\int \mathrm{d}^3\mathbf{x} \iota_0 {x}^{i}+\frac{1}{c^3}\frac{1}{10}\frac{\mathrm{d}^2}{\mathrm{d}t^2}\int \mathrm{d}^3\mathbf{x} |x|^2\iota_0{x}^{i}+\frac{1}{c^2}\frac{3}{10}\frac{\mathrm{d}}{\mathrm{d}t}\int\mathrm{d}^3\mathbf{x}\iota^k \hat{x}^{ki}+\mathcal{O}(c^{-4}),\\
    \hat{\mathrm{B}}^i=&\int \mathrm{d}^3\mathbf{x}\epsilon_{ijk} x_k\iota_j+\frac{1}{c^2}\frac{1}{10}\frac{\mathrm{d}^2}{\mathrm{d}t^2}\int \mathrm{d}^3\mathbf{x}\epsilon_{ijk}|x|^2 x_k\iota_j+\mathcal{O}(c^{-4}),\\
     \hat{\mathrm{E}}^{ij}=&\frac{1}{c}\int \mathrm{d}^3\mathbf{x} \iota_0 \hat{x}^{ij}+\frac{1}{c^3}\frac{1}{14}\frac{\mathrm{d}^2}{\mathrm{d}t^2}\int \mathrm{d}^3\mathbf{x} |x|^2\iota_0\hat{x}^{ij}+\frac{1}{c^2}\frac{5}{21}\frac{\mathrm{d}}{\mathrm{d}t}\int\mathrm{d}^3\mathbf{x}\iota_k \hat{x}^{kij}+\mathcal{O}(c^{-4}),\\
    \hat{\mathrm{I}}^{ij}=&\int \mathrm{d}^3\mathbf{x} \Sigma \hat{x}^{ij}+\frac{1}{c^2}\frac{1}{14}\frac{\mathrm{d}^2}{\mathrm{d}t^2}\int \mathrm{d}^3\mathbf{x} |x|^2\Sigma\hat{x}^{ij}-\frac{1}{c^2}\frac{20}{21}\frac{\mathrm{d}}{\mathrm{d}t}\int\mathrm{d}^3\mathbf{x}\Sigma_k \hat{x}^{kij}+\mathcal{O}(c^{-4}),
\end{align}}
  {where $\Sigma={c^{-2}} (\tau^{0}_{0}+\tau^i_i),\Sigma_i={c^{-1}}\tau_{0i},\Sigma^{ij}=\tau^{ij}$,
\begin{equation}
    \Sigma=\sum_{A=1,2} m_A\left[1+\frac{1}{c^2}\left(\frac{3}{2}v_A^2-V|_A\right)\right]\delta_A+\frac{1}{c^2}\epsilon_0\partial_k (\varphi|_1+\varphi|_2)\partial^k (\varphi|_1+\varphi|_2)+\mathcal{O}(c^{-4}),
\end{equation}
where $(\partial_k \varphi|_1)^2$ and $(\partial_k \varphi|_2)^2$ are divergent terms, which will be eliminated by the partie-finie integration. The interaction term $\partial_k \varphi|_1 \partial^k \varphi|_2$ can be calculated by
\begin{equation}
    \underset{C=0}{\mathrm{FP}} \int \mathrm{d}^3\mathbf{x} \left(\frac{r}{r_0} \right)^C\hat{x}^{ij}\partial_k \frac{1}{|\mathbf{x}-\mathbf{x}_1|}\partial^k \frac{1}{|\mathbf{x}-\mathbf{x}_2|}=\frac{2\pi}{r}(\hat{x}_1^{ij}+\hat{x}_2^{ij}).
\end{equation}
}
We expand the mass octupole and the mass current quadrupole to the Newtonian order
\begin{align}
    \hat{\mathrm{I}}^{ijk}=\int \mathrm{d}^3\mathbf{x} \Sigma \hat{x}^{ijk}+\mathcal{O}(c^{-2}),\quad \hat{\mathrm{J}}^{ij}=\int \mathrm{d}^3\mathbf{x} \epsilon_{ijk}x_{i}\Sigma_j \hat{x}^{ij}+\mathcal{O}(c^{-2}),
\end{align}
where we have used the relationships of STF tensor
\begin{gather}
x^{\langle i}v^{k\rangle}v^k=\frac{1}{2}v^2x^{i}-\frac{1}{6}r\dot{r}v^i,\quad x^{\langle ik \rangle}x^k=\frac{2}{3}r^2x^i\\
    x^{\langle ij}v^{k\rangle}v^k=\frac{1}{3}x^{\langle ij \rangle}v^2+\frac{2}{5}r^{\langle i}v^{j\rangle}r\dot{r}-\frac{2}{15}v^{\langle ij \rangle}r^2,\quad x^{\langle ijk \rangle}x^k=\frac{3}{5}x^{\langle ij\rangle}r^2.
\end{gather}

We use the acceleration in Eq. \eqref{aA} to replace $\ddot{x}_i$ when computing the time derivatives of the mass quadrupole. In the center of mass frame, we obtain the EM dipole at 1PN order 
\begin{align}
    \mathrm{E}_{i}=&\mu\xi_1x_i+\frac{\mu}{c^2}\bigg\{\bigg[-\frac{1}{2}\xi_0\frac{Gm}{r}\mathcal{Z}-\xi_2\frac{Gm}{r}-\frac{1}{10}\xi_3\frac{G m}{r}\mathcal{Z}+\frac{1}{2}\xi_3 v^2\bigg]x_i+\frac{3}{10}\xi_3 r\dot{r}v_i\bigg\}+\mathcal{O}(c^{-4}),\\
    \mathrm{E}_i^{(1)}=&\mu\xi_1v_i+\frac{\mu}{c^2}\bigg\{\bigg[\Big(-\xi_2-\frac{1}{2}\xi_0-\frac{2}{5}\xi_3\Big)\frac{G m}{r}\mathcal{Z}+\frac{4}{5}\xi_3v^2\bigg]v_i+\bigg[\xi_2+\Big(\frac{1}{2}\xi_0-\frac{6}{5}\xi_3\Big)\mathcal{Z}\bigg]\frac{Gm}{r^2}\dot{r}x_i\bigg\}+\mathcal{O}(c^{-4}),\\
    \mathrm{E}_i^{(2)}=&-\mu\xi_1\frac{G m}{r^2}\mathcal{Z}x_i+\frac{\mu}{c^2}\bigg\{\bigg[\Big(4-2\eta+(1-2\eta)\mathcal{Q}\Big)\xi_1+2\xi_2+\Big(\xi_0-\frac{12}{5}\xi_3\Big)\mathcal{Z}\bigg]\frac{Gm}{r^2}\dot{r}\notag\\
    &+\bigg[\Big(\frac{Gm}{r^2}\Big)^2\Big(4+2\eta+\zeta_0-(5+4\eta)\mathcal{Q}+2\eta\mathcal{Q}^2\Big)\xi_1+\frac{8}{5}\xi_3\Big(\frac{Gm}{r^2}\mathcal{Z}\Big)^2\notag\\
    &+\frac{Gm}{r^3}\bigg[\xi_2-\Big(\Big(1+3\eta+\frac{1}{2}\mathcal{Q}\Big)\xi_1+\mathcal{Z}\Big(\frac{1}{2}\xi_0-2\xi_3\Big)\Big)v^2+\Big(\frac{3}{2}\eta\xi_1-\frac{3}{2}\eta \mathcal{Q}\xi_1-3\xi_2-\mathcal{Z}\Big(\frac{3}{2}\xi_0-\frac{18}{5}\xi_3\Big)\Big)\dot{r}^2\bigg]\bigg]x_i\bigg\}\notag\\[4pt]
    &+\mathcal{O}(c^{-4}),
\end{align}

When $\chi_B=\chi_A$ and $q_A=q_B$, the Newtonian electric-dipole contribution vanishes, whereas higher-order EM dipole terms may remain in the 1PN order. 
The EM current dipole is 
\begin{align}
    \mathrm{B}_i=&\mu\epsilon_{ijk}x_{\langle j}v_{k\rangle}\zeta_2+\frac{\mu}{c^2}\epsilon_{ijk}x_{\langle j}v_{k\rangle}\bigg\{\zeta_4\bigg(2v^2-2\frac{Gm}{r}\mathcal{Z}\bigg)-\zeta_3\frac{Gm}{r}\bigg\}+\mathcal{O}(c^{-4}),\\
    \mathrm{B}^{(1)}_i=&\frac{\mu}{c^2}\epsilon_{ijk}x_{\langle j}v_{k\rangle}\bigg\{\bigg[\frac{1}{10}\zeta_3+\zeta_2 (-2 \eta -2 \eta  \mathcal{Q}+\mathcal{Q}+4)-\frac{1}{5} \zeta_4 \mathcal{Z}\bigg]\frac{Gm}{r^2}\dot{r}\bigg\}+\mathcal{O}(c^{-4}),\\
    \mathrm{B}^{(2)}_i=&\frac{\mu}{c^2}\epsilon_{ijk}x_{\langle j}v_{k\rangle}\bigg\{\bigg[\frac{1}{10}\zeta_3+\zeta_2 (-2 \eta -2 \eta  \mathcal{Q}+\mathcal{Q}+4)-\frac{1}{5} \zeta_4 \mathcal{Z}\bigg]\frac{Gm}{r^3}\bigg(v^2-3\dot{r}^2-\frac{Gm}{r}\mathcal{Z}\bigg)\bigg\}+\mathcal{O}(c^{-4}).
\end{align}
The Newtonian-order contributions of $\mathrm{B}^{(1)}_i$ and $\mathrm{B}^{(2)}_i$  vanish. The leading order terms of the EM current dipole in Eqs. \eqref{FEM}-\eqref{NEM}  become $\mathcal{O}(c^{-9})$, which will be neglected in our calculations. The mass quadrupole is
\begin{align}\label{DI}
    \hat{\mathrm{I}}_{ij}=&\mu x_{\langle i}x_{j\rangle}+\frac{\mu}{c^2}\bigg\{\bigg[\frac{29}{42}(1-3\eta)v^2  {-\frac{1}{7}(5-8\eta)\frac{G m}{r}\mathcal{Z}}\bigg]x_{\langle i}x_{j\rangle}-\frac{4}{7}(1-3\eta)r\dot{r}x_{\langle i}v_{j\rangle}+\frac{11}{21}(1-3\eta)r^2v_{\langle i}v_{j\rangle}\bigg\}\notag\\
    &+\mathcal{O}(c^{-4}),\\
    \hat{\mathrm{I}}^{(1)}_{jk}=&2\mu x_{\langle j} v_{k\rangle}\notag\\
    &+\frac{1}{c^2}\mu\bigg\{\bigg[\Big(-\frac{40}{21}+\frac{26}{7}\eta\Big)\frac{Gm}{r}\mathcal{Z}+\Big(\frac{17}{21}-\frac{17}{7}\eta\Big)v^2\bigg]v_{\langle j}x_{k\rangle}+\bigg[\frac{10}{21}-\frac{10}{7}\eta\bigg]r\dot{r}v_{\langle j}v_{k\rangle}-\frac{Gm}{r^2}\mathcal{Z}\bigg[\frac{2}{21}+\frac{9}{7}\eta\bigg]\dot{r}x_{\langle j}x_{k\rangle}\bigg\}\notag\\
    &+\mathcal{O}(c^{-4}),\\
    \hat{\mathrm{I}}_{jk}^{(2)}=&-2 \frac{G m}{r^3}\mathcal{Z}x_{\langle j} x_{k\rangle}+2 v_{\langle j}v_{k\rangle}\notag\\
    &+\frac{1}{c^2}\mu\bigg\{\bigg[\Big(-\frac{50}{21}+\frac{36}{7}\eta\Big)\frac{Gm}{r}\mathcal{Z}+\Big(\frac{9}{7}-\frac{27}{7}\eta\Big)v^2\bigg]v_{\langle j}v_{k\rangle}+\frac{Gm}{r^2}\bigg[8-4\eta+(2-4\eta)\mathcal{Q}+\Big(-\frac{6}{7}+\frac{46}{7}\eta\Big)\mathcal{Z}\bigg]\dot{r}v_{\langle j}x_{k\rangle}\notag\\
    &+\frac{Gm}{r^3}\bigg[\frac{Gm}{r}\Big(8+4\eta+2\zeta_0+(-10-8\eta)\mathcal{Q}+4\eta\mathcal{Q}^2+(2-5\eta)\mathcal{Z}^2\Big)\notag\\
    &+\Big(-2-6\eta-\mathcal{Q}+\Big(-\frac{19}{21}+\frac{26}{7}\eta\Big)\mathcal{Z}\Big)v^2+\Big(3\eta-3\eta\mathcal{Q}+\Big(\frac{2}{7}-\frac{27}{7}\eta\Big)\mathcal{Z}\Big)\dot{r}^2\bigg]\dot{r}x_{\langle j}x_{k\rangle}\bigg\}\notag\\
    &+\mathcal{O}(c^{-4}),\\
     \hat{\mathrm{I}}^{(3)}_{jk}=&-2 \frac{G m\mu}{r^3}\mathcal{Z}\Big(4x_{\langle j} v_{k\rangle}-3\frac{\dot{r}}{r} x_{\langle j} x_{k\rangle} \Big)\notag\\
     &+\frac{1}{c^2}\frac{G m\mu}{r^3}\bigg\{ \bigg[24-12\eta+(6-12\eta)\mathcal{Q}+\Big(-\frac{22}{21}+\frac{64}{7}\Big)\mathcal{Z}\bigg]r\dot{r}v_{\langle j}v_{k\rangle}\notag\\
     &+\bigg[\frac{Gm}{r}\Big(32+16\eta-8\zeta_0+(-40-32\eta-2\mathcal{Z}+4\eta\mathcal{Z})\mathcal{Q}+16\eta\mathcal{Q}^2+(-8+4\eta)\mathcal{Z}+\Big(\frac{202}{21}-\frac{188}{7}\Big)\mathcal{Z}^2\Big)\notag\\
     &+\Big(-28\eta+(-2-4\eta)\mathcal{Q}+\Big(-\frac{110}{21}+\frac{152}{7}\eta\Big)\mathcal{Z}\Big)v^2+\Big(-24+24\eta+6\mathcal{Q}+\Big(\frac{22}{7}-\frac{192}{7}\eta\Big)\mathcal{Z}\Big)\dot{r}^2\bigg]v_{\langle j}x_{k\rangle}\notag\\
     &+\bigg[\frac{Gm}{r^2}\Big(32+16\eta+\Big(40+32\eta-2\mathcal{Z}+10\eta\mathcal{Z}\Big)\mathcal{Q}\Big)+16\eta\mathcal{Q}^2+(-8+4\eta)\mathcal{Z}+\Big(-\frac{124}{24}+\frac{96}{7}\eta\Big)\mathcal{Z}^2\notag\\
     &+\Big(6+24\eta+(3-6\eta)\mathcal{Q}+\Big(\frac{23}{7}-\frac{132}{7}\Big)\mathcal{Z}\Big)\frac{\dot{r}}{r}v^2+\Big(-15\eta+15\mathcal{Q}+\Big(-\frac{10}{7}+\frac{135}{7}\Big)\mathcal{Z}\Big)\frac{1}{r}\dot{r}^3\bigg]x_{\langle j}x_{k\rangle}\bigg\}\notag\\
     &+\mathcal{O}(c^{-4}).
\end{align}
If we set $q_A=q_B=0$, this equation will return to the standard post-Newtonian mass quadrupole expansion \cite{PhysRevD.46.1517}. 
\begin{align}
    \hat{\mathrm{E}}_{ ij}=&\mu\zeta_2 x_{\langle i}x_{j\rangle}+\frac{\mu}{c^2}\bigg\{\bigg[-\zeta_3\frac{G m}{r}+\zeta_4\Big(-\frac{3}{7}\frac{Gm}{r}\mathcal{Z}+\frac{8}{21}v^2\Big)\bigg]x_{\langle i}x_{j\rangle}+\frac{6}{7}\zeta_4r\dot{r}x_{\langle i}v_{j\rangle}+\frac{1}{21}\zeta_4r^2v_{\langle i}v_{j\rangle}\bigg\}+\mathcal{O}(c^{-4}),\\
    \hat{\mathrm{E}}^{(1)}_{jk}=&2\mu\zeta_2 x_{\langle j} v_{k\rangle}+\frac{\mu}{c^2}\bigg\{\bigg[\frac{Gm}{r}\Big(-2\zeta_3-\frac{38}{21}\mathcal{Z}\zeta_4\Big)+\frac{34}{21}\zeta_4v^2\bigg]v_{\langle j}x_{k\rangle}+\frac{20}{21}\zeta_4r\dot{r}v_{\langle j}v_{k\rangle}+\frac{Gm}{r^2}\bigg[\zeta_3-\frac{25}{21}\mathcal{Z}\zeta_4\bigg]\dot{r}x_{\langle j}x_{k\rangle}\bigg\}\notag\\
    &+\mathcal{O}(c^{-4}),\\
    \hat{\mathrm{E}}_{jk}^{(2)}=&-\mu2 \frac{G m}{r^3}\zeta_2\mathcal{Z}x_{j} x_{k}+2\zeta_2 v_{j}v_k\notag\\
    &+\frac{\mu}{c^2}\bigg\{\frac{Gm}{r^2}\bigg[\Big(8-4\eta+2\mathcal{Q}-4\eta\mathcal{Q}\Big)\zeta_2+4\zeta_3-\frac{40}{7}\mathcal{Z}\zeta_4\bigg]v_{\langle j}x_{k\rangle}+\bigg[\frac{Gm}{r}\Big(-2\zeta_3-\frac{58}{21}\mathcal{Z}\zeta_4\Big)+\frac{18}{7}\zeta_4v^2\bigg]v_{\langle j}v_{k\rangle}\notag\\
    &+\bigg[\Big(\frac{Gm}{r^2}\Big)^2\Big(\Big(8+4\eta-4\zeta_0-(4+8\eta)\mathcal{Q}+4\eta\mathcal{Q}^2\Big)\zeta_2+\mathcal{Z}\zeta_3+3\mathcal{Z}^2\zeta_4\Big)\notag\\
    &+\frac{Gm}{r^3}\Big(\Big(-2-6\eta-\mathcal{Q}\Big)\zeta_2+\zeta_3-\frac{59}{21}\mathcal{Z}\zeta_4\Big)v^2+\frac{Gm}{r^3}\Big(\Big(3\eta-3\eta\mathcal{Q}\Big)\zeta_2-3\zeta_3+\frac{25}{7}\mathcal{Z}\zeta_4\Big)\dot{r}^2\bigg]x_{\langle j}x_{k\rangle}\bigg\}\notag\\
    &+\mathcal{O}(c^{-4}),\\
     \hat{\mathrm{E}}^{(3)}_{jk}=&-2\mu \frac{G m}{r^3}\mathcal{Z}\zeta_2\Big(4x_{\langle j} v_{k\rangle}-3\frac{\dot{r}}{r} x_{\langle j} x_{k\rangle} \Big)\notag\\
     &+\frac{\mu}{c^2}\bigg\{\frac{G m}{r^2}\bigg[\Big(24-12\eta+(6-12\eta)\mathcal{Q}\Big)\zeta_2+6\zeta_3-\frac{170}{21}\mathcal{Z}\zeta_4\bigg]\dot{r}v_{\langle j}v_{k\rangle}\notag\\
     &+\bigg[2\frac{Gm}{r^2}\Big(\Big(16+8\eta-8\zeta_0-4\mathcal{Z}+2\eta \mathcal{Z}-(8+16\eta-\mathcal{Z}+2\eta\mathcal{Z})\mathcal{Q}+8\eta\mathcal{Q}^2\Big)\zeta_2+\mathcal{Z}\zeta_3+\frac{181}{21}\mathcal{Z}^2\zeta_4\Big)\notag\\
     &+\Big(\Big(-28\eta-(2+4\eta)\mathcal{Q}+6\zeta_3-\frac{346}{21}\mathcal{Z}\zeta_4\Big)\zeta_2\Big)v^2+\Big(\Big(-24+24\eta-6\mathcal{Q}-18\zeta_3-\frac{170}{7}\mathcal{Z}\zeta_4\Big)\zeta_2\Big)\dot{r}^2\bigg]v_{\langle j}v_{k\rangle}\notag\\
     &+\bigg[2\frac{Gm}{r}\Big(\Big(-16-8\eta+8\zeta_0-2\mathcal{Z}+5\eta \mathcal{Z}-(8+16\eta+5\eta\mathcal{Z})\mathcal{Q}-8\eta\mathcal{Q}^2\Big)\zeta_2-2\mathcal{Z}\zeta_3+\frac{82}{21}\mathcal{Z}^2\zeta_4\Big)\notag\\
     &+\Big(\Big(6+24\eta+(3-3\eta)\mathcal{Q}+9\zeta_3-\frac{109}{7}\mathcal{Z}\zeta_4\Big)\zeta_2\Big)v^2-\Big(\Big(15\eta+15\eta\mathcal{Q}+15\zeta_3+\frac{125}{7}\mathcal{Z}\zeta_4\Big)\zeta_2\Big)\dot{r}^2\bigg]\frac{\dot{r}}{r^2}x_{\langle j}x_{k\rangle}\bigg\}\notag\\
     &+\mathcal{O}(c^{-4}).
\end{align}
We only need the Newtonian mass octupole as
\begin{align}
    \mathrm{I}_{ijk}=&  x_ix_j x_k-\frac{3}{5}r^2\delta_{(ij}x_{k)}+\mathcal{O}(c^{-2}),\\
    \mathrm{I}^{(1)}_{ijk}=&3  x_{(i}x_{j} v_{k)}-\frac{3}{5}r^2\delta_{(ij}v_{k)}-\frac{6}{5}r\dot{r}\delta_{(ij}x_{k)}+\mathcal{O}(c^{-2}),\\
    \mathrm{I}_{ijk}^{(2)}=&6  x_{(i} v_{j}v_{k)}-3 \frac{G m }{r^3}\mathcal{Z}x_ix_{j} x_{k}-\frac{3}{5} \delta_{(ij}\bigg[-3\frac{G m}{r}\mathcal{Z}x_{k)}+2 v^2x_{k)}+4r\dot{r}v_{k)}\bigg]+\mathcal{O}(c^{-2}),\\
     \mathrm{I}^{(3)}_{ijk}=&6  v_{i} v_{j}v_{k}- \frac{G m }{r^3}\mathcal{Z}\Big(-9\frac{\dot{r}}{r}x_ix_{j} x_{k}+21v_{(i}x_ix_{j)}\Big)-\frac{3}{5} \delta_{(ij}\bigg[-\frac{G m}{r}\mathcal{Z}\Big(7v_{k)}+5\frac{\dot{r}}{r}x_{k)}\Big)+6 v^2v_{k)}\bigg]+\mathcal{O}(c^{-2}),\\
     \mathrm{I}^{(4)}_{ijk}=&-\frac{G m }{r^3}\mathcal{Z}\bigg[-12\frac{G m }{r^3}\mathcal{Z}x_ix_{j} x_{k}+60v_{(i}v_{j} x_{k)}-90\frac{\dot{r}}{r}v_{(i}x_ix_{j)}-\frac{1}{r^2}(9v^2-45\dot{r}^2)x_ix_{j} x_{k}\bigg]\notag\\
     &+\frac{3}{5} \delta_{(ij}\frac{G m}{r}\mathcal{Z}\bigg[-12\frac{G m}{r^3}\mathcal{Z}x_{k)}+10\frac{\dot{r}}{r}v_{k)}+\frac{1}{r^2}(11v^2-15\dot{r}^2)x_{k)}\bigg]+\mathcal{O}(c^{-2}),
\end{align}
and the mass current quadrupole
\begin{align}
    \bar{\mathrm{J}}_{jk}=&  x_{(k}\epsilon_{j)lm} x_l v_m +\mathcal{O}(c^{-2}),\\
    \bar{\mathrm{J}}^{(1)}_{jk}=&  v_{(k}\epsilon_{j)lm} x_l v_m +  x_{(k}\epsilon_{j)lm} v_l v_m -\frac{G m }{r^3}\mathcal{Z} x_{(k}\epsilon_{j)lm} x_l x_m+\mathcal{O}(c^{-2}),\\
    \bar{\mathrm{J}}_{jk}^{(2)}=&-\frac{G m }{r^3}\mathcal{Z}\Big(-3\frac{\dot{r}}{r} x_l x_mx_{(k}+ 2 x_l x_m v_{(k}+ x_l v_m x_{(k}+4 v_{(l} x_{m)}x_{(k}\Big)\epsilon_{j)lm}+2  v_{(k}\epsilon_{j)lm} v_l v_m +\mathcal{O}(c^{-2}),\\
     \bar{\mathrm{J}}^{(3)}_{jk}=&-\frac{G m }{r^3}\mathcal{Z}\bigg[\Big(15\frac{\dot{r}^2}{r^2} -3\frac{v^2}{r^2}- 4 \frac{G m}{r^3}\mathcal{Z}\Big)x_l x_m x_{(k}\notag\\
     &-3\frac{\dot{r}}{r}\big(-2v_{(l} x_{m)}x_{(k}-x_{l} x_{m}v_{(k}+ x_l v_m x_{(k}\big)+ 8 v_{(l} x_{m)} v_{(k}+5v_l v_m x_{(k}+x_l v_m v_{(k}\bigg]\epsilon_{j)lm}+\mathcal{O}(c^{-2}),
\end{align}
where $\bar{\mathrm{I}}_{ijk}={\mathrm{I}}_{ijk}/(\mu\sqrt{1-4\eta})$ and $\bar{\mathrm{J}}_{jk}={\mathrm{J}}_{jk}/(\mu\sqrt{1-4\eta})$. The EM current quadrupole $\mathrm{B}_{ij}$ and octupole $\mathrm{E}_{ijk}$ in the center-of-mass frame can be expressed as
\begin{equation}
    \mathrm{B}_{ij}=\xi_3 x_{\langle i}v_{j\rangle}+\mathcal{O}(c^{-2}),\quad \mathrm{E}_{\langle ijk\rangle}=\xi_3 x_{\langle ijk\rangle}+\mathcal{O}(c^{-2}).
\end{equation}
Note that the time derivatives of $\mathrm{B}_{ij}$ and $\mathrm{E}_{\langle ijk\rangle}$ yield results analogous to those for $\mathrm{J}_{ij}$ and $\mathrm{I}_{ijk}$, with $\mu\sqrt{1-4\eta}$ replaced by $\xi_3$. By substituting these mass and electric multipoles into the equations for energy and angular momentum radiation, we can then express the resulting energy flux as follows
\begin{align}
    	\mathcal{F}_{GW}&=\frac{1}{c^5}\frac{32}{5}m\eta^2 \Big(\frac{G m}{r^2}\Big)  \biggl\{ \Big(\frac{Gm}{r}\mathcal{Z}\Big)^2\Big(v^2-\frac{11}{12}\dot{r}\Big) \notag \\
	& +\frac{1}{c^7}\frac{32}{5}m\eta^2 \Big(\frac{G m}{r^2}\Big) \Big(\frac{Gm}{r}\Big)^4 \biggl[ \frac{1}{21}(1-4\eta) \biggl] \mathcal{Z}^4 \notag\\
	&+\frac{1}{c^7}\frac{32}{5}m\eta^2 \Big(\frac{G m}{r^2}\Big) \Big(\frac{G m}{r}\Big)^3 \bigg\{  \notag\\
	&- \bigg[ \biggl( 4(\eta +2) +  4\eta  \mathcal{Q}^2 - 2(\zeta + 4 \eta -5)\mathcal{Q}\biggl)\mathcal{Z} + \frac{1}{2} \biggl( 2 \eta - 4+ (2 \eta -1) \mathcal{Q}\biggl) \mathcal{Z}^2+ \frac{1}{21}(44 - 115 \eta ) \mathcal{Z}^3\bigg]v^2\notag\\
	&+\bigg[ \biggl(  4(\eta +2) +  4\eta  \mathcal{Q}^2 - 2(\zeta + 4 \eta -5)\mathcal{Q}\biggl) \mathcal{Z} - \frac{4}{21} \biggl(  \eta +6+ (\eta +2) \mathcal{Q}\biggl) \mathcal{Z}^2- \frac{1}{21} ( 88 \eta - \frac{73}{2}) \mathcal{Z}^3\bigg]\dot{r}^2 \biggl\} \notag\\ 
	&+\frac{1}{c^7}\frac{32}{5}m\eta^2 \Big(\frac{G m}{r^2}\Big)  \Big(\frac{G m}{r}\Big)^2  \biggl\{ \notag \\
	&+ \bigg[ \biggl( 7 \eta  + \frac{1}{2} (2 \eta +1) \mathcal{Q}\biggl) \mathcal{Z}+\biggl( \frac{785}{336}- \frac{267}{28} \eta \biggl) \mathcal{Z}^2\bigg]v^4 + \bigg[ \biggl( \frac{16}{21}  \eta  + 5 - \frac{1}{4} (23 \eta -5) \mathcal{Q} \biggl) \mathcal{Z}+\bigl( \frac{127}{112} - \frac{44}{7} \eta \bigl) \mathcal{Z}^2 \bigg]\dot{r}^4\notag\\
	&+ \bigg[ \biggl( - \frac{301}{42} \eta  - \frac{929}{168} + \frac{4}{21} (32 \eta -11) \mathcal{Q}\biggl) \mathcal{Z}+\biggl( \frac{649}{42} \eta -\frac{563}{168} \biggl) \mathcal{Z}^2\bigg] v^2 \dot{r}^2 \bigg\} +\mathcal{O}(c^{-9}),\\
    \notag\\
        \mathcal{F}_{EM}&=\frac{1}{c^3}\frac{2}{3}\xi_1^2\frac{G }{r^2}\notag\\
	&+\frac{1}{c^5}\frac{2}{15}\frac{G}{r^2}\bigg\{\Big(\frac{Gm}{r}\Big)^3\bigg[10\biggl(-4- 2\eta -2\eta\mathcal{Q}^2+ \left(5- \zeta +4 \eta \right)\mathcal{Q}\biggl)\xi_1^2\mathcal{Z} -16 \xi_1\xi_3 \mathcal{Z}^3\bigg]\notag\\
	&+\Big(\frac{Gm}{r}\Big)^2\bigg[ 5\bigg[\Big(2+6 \eta+\mathcal{Q}\Big) \xi_1^2 - 2 \xi_1\xi_2 \bigg]\mathcal{Z}+ \left(12 \zeta_2^2+20 \xi_1\xi_3 -5 \eta \xi_0 \xi_1\right)\mathcal{Z}^2\bigg]v^2\notag\\
	&+\Big(\frac{Gm}{r}\Big)^2\bigg[ 5\bigg[\Big(-8 + \eta+(-2+7\eta)\mathcal{Q}\Big) \xi_1^2 + 10 \xi_1\xi_2 \bigg]\mathcal{Z}- \left(11 \zeta_2^2+12 \xi_1\xi_3 + 5 \eta \xi_0 \xi_1 \right)\mathcal{Z}^2\bigg]\dot{r}^2\bigg\}\notag\\
	&+\frac{1}{c^7}\frac{G}{r^2}\bigg\{\frac{64}{525}(1-4\eta)\zeta_3^2\Big(\frac{Gm}{r}\mathcal{Z}\Big)^4\notag\\
	&+\Big(\frac{Gm}{r}\Big)^3\bigg[-\frac{32}{5}\bigg[2 + \eta + \eta  \mathcal{Q}^2- (\zeta +2 \eta +1)\mathcal{Q}\bigg]\zeta_2^2 \mathcal{Z}-\frac{4}{5} \bigg[ \bigg( -4 + 2 \eta - (1 - 2 \eta ) \mathcal{Q}\bigg)\zeta_2^2 + \zeta_2\zeta_3\bigg]\mathcal{Z}^2\notag\\
	&\qquad\qquad\quad\;- \frac{4}{105} \bigg[181\zeta_2\zeta_4 + \frac{104}{5} (4 \eta -1) \xi_3^2\bigg]\mathcal{Z}^3\bigg]v^2\notag\\
	&+\Big(\frac{Gm}{r}\Big)^3\bigg[+ \frac{32}{5}\bigg[2 + \eta + \eta  \mathcal{Q}^2- (\zeta +2 \eta +1)\mathcal{Q}\bigg]\zeta_2^2 \mathcal{Z} + \frac{4}{15} \bigg[ \bigg( - 6 -   \eta - ( \eta +2) \mathcal{Q}\bigg)\zeta_2^2 + 4 \zeta_2\zeta_3\bigg]\mathcal{Z}^2\notag\\
	&\qquad\qquad\quad\;+ \frac{16}{105} \bigg[37 \zeta_2\zeta_4 - 6(4 \eta -1) \xi_3^2\bigg]\mathcal{Z}^3\bigg]\dot{r}^2\notag\\
	&+\Big(\frac{Gm}{r}\Big)^2\bigg[\frac{2}{5}\bigg[\left(20 +3\eta -\left(-5+23\eta\right) \mathcal{Q}\right)\zeta_2^2-8\zeta_2\zeta_3\bigg] \mathcal{Z}+ \frac{1}{70} \bigg[ 280 \zeta_2\zeta_4-171 (4 \eta -1)\xi_3^2\bigg]\mathcal{Z}^2\bigg]\dot{r}^4\notag\\
	&+\Big(\frac{Gm}{r}\Big)^2\bigg[\frac{4}{15}\bigg[\left( -33 -43\eta-\left(11-32\eta\right) \mathcal{Q}\right)\zeta_2^2+ 21\zeta_2\zeta_3\bigg] \mathcal{Z}+ \frac{1}{105}\bigg[ -1076\zeta_2\zeta_4+ 523 (4 \eta -1)\xi_3^2\bigg]\mathcal{Z}^2\bigg]v^2\dot{r}^2\notag\\
	&+\Big(\frac{Gm}{r}\Big)^2\bigg[\frac{4}{5}\bigg[\left( 14\eta+\left(2\eta+1\right) \mathcal{Q}\right)\zeta_2^2-3\zeta_2\zeta_3\bigg] \mathcal{Z}+\frac{1}{35} \bigg[\frac{6 }{7}\zeta_2\zeta_4-\frac{899}{10} (4 \eta -1)\xi_3^2\bigg]\mathcal{Z}^2\bigg]v^4\bigg\}+\mathcal{O}(c^{-9}),
    \end{align}
   
    and the angular momentum radiation as
    \begin{align}
    	\mathcal{G}^i_{GW}&=\frac{1}{c^5}\frac{16}{5}\mathcal{J}^i\eta\Big(\frac{G m}{r^2}\Big) \biggl\{  \frac{Gm}{r}\mathcal{Z} \bigg[ \frac{G m}{r}\mathcal{Z} +v^2- \frac{3}{2}\dot{r}^2\bigg]\notag\\
	&+\frac{1}{c^7}\frac{16}{5}\mathcal{J}^i\eta\Big(\frac{G m}{r^2}\Big) \Big( \frac{Gm}{r} \Big)^3 \bigg\{  \notag \\   
	&+ \bigg[ \biggl( -8 - 4\eta - 4\eta  \mathcal{Q}^2+ ( -2\zeta  + 8 \eta +10)\mathcal{Q}\biggl)\mathcal{Z}+ \biggl( 1 - \frac{1}{2} \eta +( \frac{1}{4}- \frac{1}{2} \eta ) \mathcal{Q}\biggl) \mathcal{Z}^2+\frac{1}{21}\biggl( 95 \eta - \frac{157}{4} \biggl) \mathcal{Z}^3\bigg]   \biggl\}     \notag \\
	&+\frac{1}{c^7}\frac{16}{5}\mathcal{J}^i\eta\Big(\frac{G m}{r^2}\Big)\Big(\frac{Gm}{r}\Big)^2 \biggl\{  \notag \\
	&+ \bigg[\biggl( -4- 2\eta - 2\eta  \mathcal{Q}^2+ ( -\zeta + 4\eta +5)\mathcal{Q}\biggl) + \biggl( 6\eta +1 \mathcal{Q}+2\biggl) \mathcal{Z}+ \frac{1}{21}\biggl(13 - \frac{263}{2} \eta \biggl)\mathcal{Z}^2\bigg]v^2\notag \\
	&+ \bigg[ \biggl( + 8 +4 \eta +4 \eta  \mathcal{Q}^2+ (2 \zeta -8 \eta -10)\mathcal{Q}\biggl) + \biggl( -\frac{5}{2} \eta -(\frac{3}{2} \eta -\frac{7}{2}) \mathcal{Q}-5 \biggl) \mathcal{Z}+\frac{1}{21}\biggl(\frac{71}{4} \eta + 30 \biggl) \mathcal{Z}^2\bigg]\dot{r}^2 \biggl\} \notag \\	
	&+ \frac{1}{c^7}\frac{16}{5}\mathcal{J}^i\eta\Big(\frac{G m}{r^2}\Big) \frac{G m}{r} \biggl\{ \notag \\
	& + \bigg[\frac{7}{2} \eta +(\frac{1}{4} \eta + \frac{1}{2}) \mathcal{Q}+ \frac{1}{21}\biggl( \frac{307}{8}-142 \eta \biggl) \mathcal{Z}\bigg]v^4 +\frac{G m}{r} \bigg[\frac{15}{4} \eta -\frac{15}{4} \eta  \mathcal{Q}+\biggl( \frac{95}{56} -\frac{285}{28} \eta \biggl) \mathcal{Z}\bigg] \dot{r}^4 \notag \\
	&+\bigg[\frac{3}{2}- 9 \eta +\frac{3}{2} \eta  \mathcal{Q}+\biggl(\frac{529}{28} \eta -\frac{29}{7}\biggl) \mathcal{Z}\bigg]v^2 \dot{r}^2 \bigg\}  \biggl\} +\mathcal{O}(c^{-9}),  \\
    \notag\\
		\mathcal{G}_{EM}^{i} & =  \frac{1}{c^3} \frac{2}{3} \mathcal{J}^i \xi_1^2 \Big(  \frac{Gm}{r}  \mathcal{Z}  \Big)  \Big( \frac{G }{r^2} \Big)  \notag\\
		& + \frac{1}{c^5}   \frac{G}{r^2} \mathcal{J}^i  \biggl\{    \Big(\frac{Gm}{r^2} \Big)^2  \biggl[   \biggl(-4- 2\eta -2\eta\mathcal{Q}^2+ \left(5- \zeta +4 \eta \right)\mathcal{Q}\biggl)\xi_1^2 -\frac{2}{3} \xi_1 \xi_2 \mathcal{Z}  + \Big( \frac{4}{5} \zeta_2^2  -  \frac{1}{3} \eta \xi_0 \xi_1 -  \frac{4}{3} \xi_1 \xi_3	\Big) \biggl] \notag\\
		& + \Big(   \frac{Gm}{r} \Big) \biggl[  \frac{1}{3} \biggl( 2 + 6\eta + \mathcal{Q} \biggl) \xi_1^2 
		- \frac{2}{3} \xi_1 \xi_2 
		+ \biggl( \frac{4}{5} \zeta_2^2 - \frac{1}{3} \eta \xi_0 \xi_1 + \frac{28}{15} \xi_1 \xi_3 \biggl) \mathcal{Z} \biggl] v^2 \notag\\
		& + \Big(   \frac{Gm}{r} \Big)  \biggl[  \biggl(-1 + \mathcal{Q} \biggl)  \eta \xi_1^2 + 2 \xi_2 \xi_3
		+ \biggl(   -\frac{6}{5} \zeta_2^2 + \eta \xi_0 \xi_1 - \frac{9}{5} \xi_1 \xi_3 \biggl) \mathcal{Z}\biggl] \dot{r}^2	\biggl\}	\notag\\
		& + \frac{1}{c^7} \frac{G }{r^2}  \mathcal{J}^i\biggl\{  \Big(   \frac{Gm}{r} \Big) ^3   \biggl[ -  \frac{16}{5} \biggl( 2 + \eta + \eta \Big( \mathcal{Q}\Big)^2 - (1 + \zeta + 2\eta) \mathcal{Q}  \biggl)  \zeta_2^2  \biggl]  \mathcal{Z}  \notag\\
		& + \Big( \frac{Gm}{r} \Big)^3 \biggl[ -\frac{1}{5}   \biggl(   -4 + 2\eta  - \mathcal{Q} + 2 \eta \mathcal{Q}\biggl)  \zeta^2_2  + \frac{3}{5} \zeta_2 \zeta_3 \biggl] \mathcal{Z}^2  + \Big( \frac{Gm}{r} \Big)^3  \biggl[  \frac{307}{105} \zeta_2 \zeta_4 + \frac{32}{75} (1 - 4\eta) \zeta_3^2\biggl] \mathcal{Z}^3  \notag\\
		& + \Big( \frac{Gm}{r} \Big)^2  \biggl[ -\frac{8}{5} \biggl(  2 + \eta  + \Big( \mathcal{Q} \Big)^2  \eta - (1 + \zeta + 2\eta) \mathcal{Q}\biggl) \zeta_2^2  +  \biggl[  \frac{4}{5} \biggl( 2 + 6\eta + \mathcal{Q}  \biggl) \zeta_2^2 - 2 \zeta_2 \zeta_3 \biggl] \mathcal{Z} \notag\\
		& \; \; \; \; \; \; \; \; \; \; \; \; \;\; \; \; \; \; +  \frac{1}{35}  \biggl(   2 \zeta_2 \zeta_4 + \frac{2497}{30} (-1 + 4\eta ) \xi_3^2\biggl)  \mathcal{Z}^2  \biggl] v^2 	 \notag\\
		& + \Big( \frac{Gm}{r} \Big)^2 \biggl[   
		+ \frac{16}{5} \biggl(   2 + \eta +   \Big( \mathcal{Q} \Big)^2  \eta - (1 + \zeta + 2\eta )  \mathcal{Q}  \biggl) \zeta_2^2  + \biggl[ \frac{2}{5} \biggl(   -5 (2 + \eta) + (-3 + 7\eta) \mathcal{Q} \biggl) \zeta_2^2  + \frac{13}{5} \zeta_2 \zeta_3 \biggl] \mathcal{Z}  \notag\\
		& \; \; \; \; \; \; \; \; \; \; \; \; \;\; \; \; \; \; +  \frac{1}{105} \biggl(  283 \zeta_2 \zeta_4 + \frac{493}{2} (1 - 4 \eta) \zeta_3^2\biggl)
		\mathcal{Z} ^2 \biggl] \dot{r}^2  \notag\\
		&  +  \Big( \frac{Gm}{r} \Big) \biggl[  +  \frac{1}{5} \biggl(  14 \eta  +  \mathcal{Q}  + 2\eta \Big( \frac{\mathcal{Q} }{\eta} \Big) \biggl) \zeta_2^2 - \frac{3}{5} \zeta_2 \zeta_3  + \frac{1}{105} \biggl( 281 \zeta_2 \zeta_4 + \frac{356}{5} (-1 + 4\eta ) \zeta_3^2 \biggl) \mathcal{Z} \biggl] v^4 \notag\\
		& +  \Big( \frac{Gm}{r} \Big) \biggl[  -  3\biggl( -1 + \mathcal{Q} \biggl) \eta  \zeta_2^2 - 3 \zeta_2 \zeta_3 
		+ \biggl(  \frac{25}{7}  \zeta_2 \zeta_4  - \frac{12}{7} (1 - 4 \eta) \zeta_3^2\biggl) \mathcal{Z}\biggl] \dot{r}^4  \notag\\
		& + \Big( \frac{Gm}{r} \Big) \biggl[  \frac{42}{35} \biggl(  1 +  ( -6 + \mathcal{Q} ) \eta  \biggl) \zeta_2^2 + \frac{126}{35}  \zeta_2 \zeta_3  + \frac{2}{35} \biggl( -42 \zeta_2 \zeta_4 + 44 (1 - 4\eta) \zeta_3^2 \biggl) \mathcal{Z} \biggl] v^2 \dot{r}^2  \biggl\} + \mathcal{O}(c^{-9}).
\end{align}
\end{widetext}
The GW and EM energy and angular-momentum fluxes above are valid for generic inspiralling binary orbits. We have simplified them by applying to the circular orbits in Sec. \ref{Sec4}.

\section{The innermost stable circular orbit of PN theory and RN metric}\label{App.C}

In this section, we will check the consistency of the ISCO in the RN metric in Eq. \eqref{RN} \cite{Chandrasekhar} and in the PN frame. The RN spacetime admits two Killing vectors, $k=\partial_t$ and  $\chi=\partial_\phi$ corresponding to two conserved quantities, the orbital energy and the angular momentum
\begin{align}
    \frac{E}{m_0}&=-k_\mu u^\mu=f(r)\dot{t},\\
    \frac{L}{m_0}&=\chi_\mu u^\mu=r^2\dot{\phi},
\end{align}
where $u^\mu$ is the four-velocity, $\dot{A}$ is $\mathrm{d}A/\mathrm{d}\tau$ and $\tau$ is the proper time, $m_0$ is the mass of the test particle and denotes $\bar{E}=E/m_0$ and $\bar{L}=L/m_0$. Using $g_{\mu\nu}u^\mu u^\nu=-c^2$, we obtain the effective potential $V_{eff}(r)$ as
\begin{align}\label{RNE2}
    \bar{E}^2=&\dot{r}^2+V_{eff}(r),\\
    V_{eff}(r)=&f(r)\Big(c^2+\frac{\bar{L}^2}{r^2}\Big)
\end{align}
For a circular orbit, we have $\partial_r V_{eff}=0$ and $\dot{r}=0$ in Eq. \eqref{RNE2}, solving these two equations gives the energy $\bar{E}$ and the angular momentum $\bar{L}$ as
\begin{align}
    \bar{E}^2=\frac{f(r)^2}{1-3M/r+2{Q}^2/r^2},\\
    \bar{L}^2=\frac{r(M-{Q}^2/r)}{1-3M/r+2{Q}^2/r^2},
\end{align}
where ${Q}$ is the charge of the RN metric. Here we obtain Kepler's third law by the definition of angular frequency
\begin{align}
    \omega=\frac{\mathrm{d}\phi}{\mathrm{d}t}=\frac{\dot{\phi}}{\dot{t}}=\frac{f(r)}{r^2}\frac{\bar{L}}{\bar{E}},
\end{align}
and 
\begin{align}
    \omega^2=\frac{Gm}{r^3}\bigg(1-\frac{GM}{c^2r}\frac{{Q}^2}{M^2}\bigg)+\mathcal{O}(c^{-4}),
\end{align}
which is similar to the PN result in Eq. \eqref{omega} if we set $q_A\neq 0$, $q_B= 0$, and $m_B\ll m_A$, but not the same as Eq. \eqref{omega}.

For a stable circular orbit, we have $\partial_r^2 V_{eff}(r)>0$ and obtain an equation for the RN ISCO radius $r^{(RN)}_{ISCO}$ while $\partial_r^2 V_{eff}(r)=0$
\begin{equation}
    (r^{(RN)}_{ISCO})^3-6\frac{G}{c^2}M (r^{(RN)}_{ISCO})^2+9\frac{G}{c^2}{{Q}}^2r^{(RN)}_{ISCO}-4\frac{G}{c^2}\frac{{{Q}}^4}{M}=0.
\end{equation}
The solution of this equation is
\begin{align}
    r^{(RN)}_{ISCO}&=3\frac{G M}{c^2}+\sqrt{9\frac{G^2 M^2}{c^4}-8\frac{G^2 {Q}^2}{c^4}}\notag\\
    &=6\frac{G M}{c^2}-\frac{4}{3}\frac{G {Q}^2}{c^2M}+\mathcal{O}\bigg[\Big(\frac{{Q}}{M}\Big)^4\bigg],
\end{align}

where we have expanded the small charge. This ISCO radius does not agree with our result in Eq. \eqref {rISCO} in the PN frame. This is because they are in the Schwarzschild and harmonic coordinates, respectively. So we need to compare the frequency solution in the RN metric with that in the PN frame. Substituting Kepler's third law into $r^{(RN)}_{ISCO}$, we obtain
\begin{align}
    x^{(RN)}_{ISCO}=\frac{1}{6}\Big(1+\frac{1}{6}\frac{{Q}^2}{M^2}\Big)+\mathcal{O}\bigg[\Big(\frac{{Q}}{M}\Big)^4\bigg].
\end{align}

We finally obtain the same result as Eq. \eqref{OmegaISCO}, by setting $q_A\neq 0$, $q_B= 0$, and $m_B\ll m_A$ and treating $\zeta_0$ as a small quantity.

\bibliography{BIB.bib}
\bibliographystyle{apsrev4-1}

\end{document}